\documentclass[preprint,nofootinbib,floatfix]{revtex4-2}
\usepackage{amsmath,amssymb,amsfonts,color,graphicx,graphics,latexsym,placeins,epsfig,multirow}
\usepackage[a4paper, total={6.8in, 10in}]{geometry}
\usepackage{graphicx}
\usepackage{epsfig}
\usepackage{epstopdf}
\usepackage{amsfonts}
\usepackage{amssymb}
\usepackage{amsbsy}
\usepackage{amsmath}
\usepackage{mathrsfs}
\usepackage{latexsym}
\usepackage{natbib}
\usepackage{bm}
\usepackage{color}
\usepackage{braket}
\usepackage{slashed}
\usepackage{pgfplots}
\usepackage{comment}
\usepackage[normalem]{ulem}

\usepackage{tikz}
\usetikzlibrary{shapes,arrows,shadows}
\usepackage{hyperref}
\hypersetup{
	colorlinks=true,       % false: boxed links; true: colored links
	linkcolor=blue,          % color of internal links
	citecolor=blue,        % color of links to bibliography
	filecolor=blue,      % color of file links
	urlcolor=blue           % color of external links
}

\def\beq{\begin{equation}}
\def\eeq{\end{equation}}
\def\br{\begin{eqnarray}}
\def\er{\end{eqnarray}}

\begin{document}

\author{Susobhan Mandal}
\email{sm12ms085@gmail.com}

\author{S. Shankaranarayanan}
\email{shanki@iitb.ac.in}

\affiliation{ Department of Physics, 
Indian Institute of Technology Bombay,
Mumbai - 400076, India }


\date{\today}

\title{Dynamical 4-D Gauss-Bonnet action from matter-graviton interaction at one-loop}

\begin{abstract}
The occurrence of singularities at the centers of black holes suggests that general relativity (GR), although a highly successful model of gravity and cosmology, is inapplicable. This is due to the breakdown of the equivalence principle. Gauss-Bonnet (GB) action is the simplest extension of GR as it possesses second-order equations of motion and is devoid of ghosts. However, in 4-D, the GB action is topological. Recently, Glavan and Lin proposed a mathematical framework that transforms the 4-D GB gravity theory into a non-topological one. However, it has been argued that without a canonical way to choose 4-D from the higher-dimensional space, such a GB gravity is not well-defined in 4-D. Naturally, there has been much interest in having a systematic procedure for making the 4-D GB term non-topological, such as using the counterterm regularization method in 4-D, regularization with the dimensional derivative, and Kaluza–Klein reduction. The current work takes a step in addressing this issue by demonstrating that the rescaling of the GB coupling $\alpha \rightarrow \alpha/(D - 4)$ arises from the self-energy correction of gravitons in 4-D using \emph{only} the established quantum field theoretic techniques.
%via \emph{the dimensional regularization}. 
To keep things transparent, we focus on the linearized theory of gravity coupled with matter fields. By computing the one-loop self-energy correction of gravitons induced by the matter fields, we explicitly provide the origin of the prescription provided by Glavan and Lin. We compare the procedure with other regularization procedures like Kaluza-Klein dimensional reduction and conformal scaling regarding the strong coupling problem. Our work naturally opens a new window to considering 4-D Einstein Gauss-Bonnet gravity as the most straightforward modification to GR.

\end{abstract}

\maketitle

\pagebreak
\newpage

\section{Introduction}
General relativity (GR) is a highly successful physical theory,  arising as a natural extension of special relativity. Over the years, it has yielded numerous accurate experimental validations of its core predictions~\cite{Will:2014kxa}. Fundamentally,  GR operates as a geometric theory, positing that the gravitational field influences spacetime geometry. This concept allows for developing a theory where geometric properties like metric and connection become dynamic variables, enabling a non-trivial description of spacetime in terms of curvature.

The occurrence of singularities at the centers of black holes raises doubts about the universality of 
GR~\cite{2009-Alexander.Yunes-PRep,2010-DeFelice.Tsujikawa-LivingRev.Rel.,2010-Sotiriou.Faraoni-Rev.Mod.Phys.,2011-Capozziello.DeLaurentis-Phys.Rept.,2011-Hinterbichler-RMP,2012-Clifton.etal-Phys.Rept.,2014-deRham-LRR,2015-Joyce.etal-Phys.Rept.,2017-Nojiri.etal-Phys.Rept.,2022-Shanki.Joseph-GRG}. Paradoxically, the very success of GR prompts the need for modifications \cite{jackiw2003chern, alexander2009chern, zhao2010probing, heisenberg2019systematic, carballo2018minimally, kreienbuehl2012modified, donoghue1994general}. On the largest scale, a surprising revelation from observational cosmology is that the current Universe is accelerating. 
This can possibly be explained by the presence of an exotic matter source called dark energy or by various modifications to GR on the largest length scales \cite{bloomfield2013dark, kunz2007dark, bahamonde2018dynamical, wei2008distinguish, joyce2016dark, capozziello2010dark, battye2012effective, afshordi2007cuscuton, tsujikawa2010modified, cusin2018dark, navarro2006modified, abdalla2005consistent, borowiec2007dark, tsujikawa2007matter}. Theoretically, a significant challenge lies in the quantum description of gravity. In $D-$dimensional spacetime, the gravitational constant ($G_D$) has a negative mass dimension ($2 - D$). Consequently, Einstein-Hilbert gravity is non-renormalizable, necessitating infinite counterterms for a consistent description \cite{deser1957general, donoghue1997perturbative, odintsov1992general, klauder1975meaning, donoghue1995introduction, shomer2007pedagogical}.
There are many ways of modifying GR in the small scales (strong gravity) and cosmological scales~\cite{2022-Shanki.Joseph-GRG}. In this context, Lovelock's theorem plays a crucial role as a guiding principle for classifying modified gravity theories, offering a robust framework for their exploration~\cite{lovelock1971einstein, lovelock1972four, lanczos1938remarkable,Padmanabhan:2013xyr}. Lovelock's theorem states that the Einstein Field equations are the only possible solution to a classical action that contains up to second derivatives of the 4-D spacetime metric $(g_{\mu\nu})$.   However, a significant shift occurred in 2020 when Glavan and Lin introduced singular regularization of the Gauss-Bonnet (GB) term in 4-D, leading to non-trivial equations of 
motion~\cite{glavan2020einstein} (for earlier works, see \cite{tomozawa2011quantum, PhysRevD.88.024006}). More specifically,  Glavan and Lin showed that if one scales the GB  coupling by $\alpha \rightarrow \alpha/(D - 4)$ and then considers 
the limit $D \rightarrow 4$, it turns out that now there is no obstacle in considering the  GB term on the same level as the Einstein-Hilbert term. 
In that sense,  Glavan and Lin's procedure is an extension of the dimensional regularization of the Ricci scalar in 2-D~\cite{1993-Mann.Ross-CQG}.

4-D GB gravity exhibits interesting phenomena in black holes, cosmology, and weak-field gravity~\cite{glavan2020einstein,fernandes20224d, konoplya2020stability, konoplya2020black, Konoplya:2020bxa, wang2021hawking, ovgun2021black, chatterjee2014second, wei2021testing, ghosh2020generating, konoplya2020grey, islam2020gravitational, ghosh2020radiating, ai2020note, kumar2020gravitational, wang20214d, aoki2021inflationary, feng2021theoretical, clifton2020observational, sadjadi2020cosmic, aoki2020cosmology, mishra2020quasinormal, bonifacio2020amplitudes}. However, there has been criticism of Glavan and Lin's procedure of considering the GB gravity in 4-D as a set of solutions of higher-dimensional GB gravity~\cite{Ai:2020peo,Gurses:2020ofy,Hennigar:2020lsl,Lu:2020iav,Shu:2020cjw, mahapatra2020note}. More specifically, it was shown that the naive limit of the higher-dimensional theory to 4-D needs to be better defined by considering simple spacetimes beyond spherical symmetry, such as Taub-NUT spaces, and contrasting the resultant metrics with the actual solutions of the new theory. 
Also, it was shown that the naive application of $\alpha \rightarrow \alpha/(D - 4)$ leads to divergence of black hole entropy~\cite{Lu:2020iav}. It has been argued that without a canonical way to choose 4-D from the higher-dimensional space, such a GB gravity is not well-defined in 4-D~\cite{fernandes20224d}. 

Naturally, there has been much interest in having a systematic procedure for making the 4-D GB term non-topological, such as using the counterterm regularization method in 4-D, regularization with the dimensional derivative, and Kaluza–Klein 
reduction (for the details, see the review \cite{fernandes20224d}). While these procedures provide novel ways to understand the dynamical origin of 4-D GB gravity physically, they also lead to strong coupling problem ~\cite{fernandes20224d, Lu:2020iav, Kobayashi:2020wqy, Fernandes:2020nbq, Hennigar:2020lsl, Fernandes:2021dsb, Aoki:2020lig, Clifton:2020xhc, Bonifacio:2020vbk}. In the leading order, all these approaches lead to the GB term. However, they also contain subleading terms that are, in principle, different. Many times, such contributions are divergent in nature \cite{Lu:2020iav, Kobayashi:2020wqy, Fernandes:2020nbq, Hennigar:2020lsl, Fernandes:2021dsb, Aoki:2020lig, Clifton:2020xhc, Bonifacio:2020vbk}. As a result, the required counterterms to cancel the divergent terms to get finite physical observables depend on the regularisation procedure.

This leads us to the following question:  Can we explain the dynamical origin of the GB term in 4-D using only the established quantum field theoretic techniques? If yes, can this approach avoid the strong coupling problem mentioned above? What are the subleading terms, and can those subleading terms be regularized? This work is a step in this direction by showing that the rescaling $\alpha \rightarrow \alpha/(D - 4)$ can be obtained through the self-energy correction of gravitons in 4-D via the dimensional regularization. Restricting the quadratic part of linearised gravity theory around the Minkowski metric, we show that the effect of matter at one-loop generates a term related to the non-topological GB gravity. We also show that the interaction generates a counterterm that must be added to the tree-level action to cancel the diverging term. We show this explicitly for massless, canonical scalar fields and massless Dirac fermions. Note that 1-loop  matter corrections of the graviton have been studied earlier~\cite{Barth:1983hb,Simon:1990ic,Buchbinder:1992rb}. To our knowledge, such an analysis is lacking in the literature and is crucial for the phenomenological studies of 4-D GB gravity~\cite{Nenmeli:2021orl,Das:2022hjp,Chowdhury:2022ktf,Mandal:2023kpu}.

\noindent \section{Effect of interaction between Gravitons and massless scalar at 1-loop}
%\label{sec:1loop-Scalar}
%
The action of a massless scalar field theory minimally coupled to gravity in $D-$dimensional space-time is:
\begin{equation}\label{action for scalar 1}
S = \frac{1}{2\kappa^{2}}\int d^{D}x \ \sqrt{-g}R - \frac{1}{2}\int d^{D}x \sqrt{-g}
g^{\mu\nu}\partial_{\mu}\phi\partial_{\nu}\phi,
\end{equation}
where $\phi$ is a massless scalar field, $\kappa^{2} = 8\pi G_D$ in natural units ($\hbar = c = 1$), and $G_D$ is the D-dimensional Newton's constant. To keep things transparent, let us consider the first-order perturbations $h_{\mu\nu}$ about the Minkowski background: $g_{\mu\nu} = \eta_{\mu\nu} 
+ \kappa h_{\mu\nu}$. Note that the metric perturbation $h_{\mu\nu}$ has mass dimension $1$ in 4-D. The Einstein-Hilbert action up to second-order in metric perturbation $h_{\mu\nu}$ is derived in the Appendices \ref{App:Linearised gravity} and \ref{App:Linearised gravity EH} for completeness. 

\begin{figure}
\includegraphics[height = 2cm, width = 5cm]{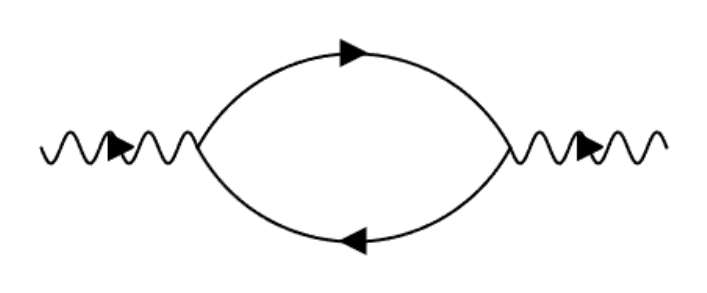}
\caption{One-loop Feynman diagram describing the self-energy correction to graviton propagator 
due to matter field}
\label{fig:self-energy1}
\end{figure}

Expanding the matter action about the Minkowski metric up to linear order leads to:
\begin{equation}
S_{M}[g,\phi] = S_{M}[\eta, \phi] + \kappa\int d^{D}x\left(\frac{1}
{\sqrt{-g}}\frac{\delta S_{M}}{\delta g_{\mu\nu}}\right)_{\eta}h_{\mu\nu},
\end{equation} 
where,
\begin{equation}
T^{\mu\nu} = \left(\frac{2}{\sqrt{-g}}\frac{\delta S_{M}}{\delta g_{\mu\nu}}\right)_{\eta}
= \partial^{\mu}\phi\partial^{\nu}\phi - \frac{1}{2}\eta^{\mu\nu}\eta^{\rho\sigma}\partial
_{\rho}\phi\partial_{\sigma}\phi,
\end{equation}
is the energy-momentum tensor of the scalar field theory. Therefore, the matter action up to linear order in metric perturbation can be expressed as:
\begin{equation}
S_{M} = - \frac{1}{2}\int d^{D}x \partial_{\mu}\phi\partial^{\mu}\phi + \kappa\int d^{D}x
h_{\mu\nu}T^{\mu\nu},
\end{equation}
where the indices are contracted \textit{w.r.t} the Minkowski metric $\eta_{\mu\nu} = \text
{diag}(-1,1,1,\ldots,1)$. Here, we adjusted the second term in the matter action by doubling it, as we similarly scaled the linearized Einstein-Hilbert action to derive the Fierz-Pauli action. Therefore, at the one-loop level, the self-energy correction to the gravitons
due to the matter field can be described by the Feynman diagram shown in Fig.~\eqref{fig:self-energy1}.
The vertex associated with the interaction between graviton and scalar field in momentum space
is expressed as
\begin{equation}
\begin{split}
\mathcal{V}_{\mu\nu}(p, p + k, - k) & = - \frac{\kappa}{2}[(p + k)_{\mu}k_{\nu} + (p + k)_{\nu}
k_{\mu}\\
 & - \eta_{\mu\nu} (p + k)^{\sigma} k_{\sigma}] \equiv \mathcal{P}_{\mu\nu}^{ \ \ \rho\sigma}(p + k)_{\rho}k_{\sigma},
\end{split}
\end{equation}
where
\begin{equation}
\mathcal{P}_{\mu\nu}^{ \ \ \rho\sigma} = - \frac{\kappa}{2}\left(\delta_{\mu}^{\rho}\delta_{\nu}
^{\sigma} + \delta_{\mu}^{\sigma}\delta_{\nu}^{\rho} - \eta_{\mu\nu}\eta^{\rho\sigma}\right).
\end{equation}
From the free part of the scalar field, we may note that the scalar field propagator is $- i/p^{2}$ where $p^{\mu}$ is  4-momentum. To evaluate the self-energy corrections to gravitons, we evaluate the following integral corresponding to the 
above Feynman diagram:
\begin{equation}
\begin{split}
\mathcal{I} & = - \int\frac{d^{D}k}{(2\pi)^{D}}\frac{1}{(p + k)^{2}k^{2}}\mathcal{P}_{\mu\nu}
^{ \ \ \rho\sigma}\mathcal{P}_{\alpha\beta}^{ \ \ \gamma\delta}(p + k)_{\rho}k_{\sigma}(p + k)
_{\gamma}k_{\delta}\\
 & = - \mathcal{P}_{\mu\nu}^{ \ \ \rho\sigma}\mathcal{P}_{\alpha\beta}^{ \ \ \gamma\delta}
 \int\frac{d^{D}k}{(2\pi)^{D}}\int_{0}^{1}\frac{dx}{[(k + px)^{2} + p^{2}x(1 - x)]^{2}}\\
 & \times [p_{\rho}k_{\sigma} + k_{\rho}k_{\sigma}][p_{\gamma}k_{\delta} + k_{\gamma}k_{\delta}],
\end{split}
\end{equation}
where in the last line, we have used Feynman's parametrization. Using the odd and even symmetries, the above integral reduces to:
\begin{equation}
\label{eq:1loppintegrals-Scalar}
\begin{split}
\mathcal{I} & = - \mathcal{P}_{\mu\nu}^{ \ \ \rho\sigma}\mathcal{P}_{\alpha\beta}^{ \ \ \gamma
\delta}\int\frac{d^{D}k}{(2\pi)^{D}}\int_{0}^{1}\frac{dx}{[(k + px)^{2} + p^{2}x(1 - x)]^{2}}\\
 & \times\Big\{p_{\rho}p_{\gamma}k_{\sigma}k_{\delta}(1 - 2x)^{2} + p_{\rho}p_{\sigma}p_{\gamma}
 p_{\delta}x^{2}(1 - x)^{2}\\
 & + p_{\rho}p_{\sigma}k_{\gamma}k_{\delta}x(x - 1) + k_{\rho}k_{\sigma}p_{\gamma}p_{\delta}
 x(x - 1) + k_{\rho}k_{\sigma}k_{\gamma}k_{\delta}\Big\}\\
 & \equiv - \mathcal{P}_{\mu\nu}^{ \ \ \rho\sigma}\mathcal{P}_{\alpha\beta}^{ \ \ \gamma\delta}
 [I_{\rho\sigma\gamma\delta}^{(1)} + I_{\rho\sigma\gamma\delta}^{(2)} + I_{\rho\sigma\gamma\delta}
^{(3)} + I_{\rho\sigma\gamma\delta}^{(4)} + I_{\rho\sigma\gamma\delta}^{(5)}]. 
\end{split}
\end{equation}
We can evaluate the above integrals for $D  = 4$. As shown in Appendix \ref{1-Loop integrals}, we can explicitly evaluate up to $1/\epsilon$ order (where $\epsilon = 4 - D$).

Hence, in $D \rightarrow 4$ dimensions, the one-loop correction to the quantum effective action due to the scalar field is
\begin{equation}
\label{eq:Scorr}
S_{\rm corr}^{(2)} = - \frac{\alpha}{4 \epsilon}\int d^{4}x 
\left[8 \kappa^2  h^{\alpha\beta}
(\Box)^{2} h_{\alpha\beta} + 4 \kappa^2  h (\Box)^{2} h \right] \, ,
\end{equation} 
where 
\begin{equation}
\alpha = \frac{1}{120(4\pi)^{2}}
\label{def:alpha}
\end{equation}

The above self-energy correction to graviton modifies the kinetic energy part of the effective action. This modification contains the operator $(\Box)^{2}$, which could only be present if one considers a quadratic theory of gravity. This is shown explicitly in the Appendix \ref{App:Linearised gravity QG}.

We are now in a position to compare the above expression with the expression with the general quadratic gravity action: 
\begin{equation}
\label{eq:GenQuadaction}
S_{\rm quad} = \int d^4x \sqrt{-g} \left[ a  R_{\mu\nu\rho\sigma}R^{\mu\nu\rho\sigma} + bR_{\mu\nu}R^{\mu\nu} + cR^{2} \right] \, ,
\end{equation}
where $a, b$ and $c$ are coupling constants. Linearising the above action w.r.t the Minkowski background, as shown in the Appendix \ref{App:Linearised gravity QG} and comparing with Eq.~\eqref{eq:Scorr}, we 
obtain the following relations:
\begin{equation}
4a + b = - \frac{8 \alpha }{\epsilon}, \ 
c - a = - \frac{4 \alpha }{\epsilon},
\end{equation}
Redefining the coupling constants as:
\begin{equation}
a = - \frac{\alpha }{\epsilon}\bar{a}, \ 
b = - \frac{\alpha }{\epsilon}\bar{b}, \ 
c = - \frac{\alpha }{\epsilon}\bar{c}, 
\end{equation}
we obtain the following relations
\begin{equation}
4\bar{a} + \bar{b} = 8, \ \bar{c} - \bar{a} = 4.
\end{equation}
Since we have two equations and three unknowns, we may write the coefficients $\bar{b}$ and $\bar{c}$ in terms of the unknown coefficient $\bar{a}$ which determines the strength of the GB term. Rewriting $\bar{b}$ and $\bar{c}$ in-terms of $\bar{a}$ we can write the quantum correction to the effective action \eqref{eq:Scorr} as follows
\begin{eqnarray}\label{loop corrected action}
S_{corr}^{(2)} & = &- \frac{\alpha}{\epsilon}\bar{a}\int d^{4}x\sqrt{-g}\Big[R_{\mu\nu\rho\sigma}R^{\mu\nu\rho\sigma} - 4R_{\mu\nu}R^{\mu\nu} + R^{2}\Big] \nonumber \\
 & -&  \frac{\alpha}{\epsilon} \int d^{4}x\sqrt{-g}\Big[8R_{\mu\nu}R^{\mu\nu} + 4R^{2}\Big] \, . 
\end{eqnarray}
This is the first key result regarding which we want to discuss the following points: First, the term inside the first integral in the RHS of the above expression is the GB term and is proportional to $(D-4)$~\cite{Padmanabhan:2013xyr}. Since the prefactor is proportional to $1/\epsilon = 1/(4 - D)$, $(D-4)$ factor that makes the GB term topological in 4-D cancels, leading to a dynamical 4-D GB term. 
To see this explicitly, we start with the Einstein-Gauss-Bonnet action without matter field described by the following action
\begin{equation}
S_{EGB} = \int d^{4}x\sqrt{-g}[R - 2\Lambda + \hat{\alpha}\mathcal{G}], \ \mathcal{G} = R_{\mu\nu\rho\sigma}R^{\mu\nu\rho\sigma} - 4R_{\mu\nu}R^{\mu\nu} + R^{2},
\end{equation}
where $\hat{\alpha} = \alpha/(D - 4)$ and $\alpha$ is the strength of the Gauss-Bonnet coupling, then the field equations can be expressed as
\begin{equation}\label{modified Einstein's equations}
G_{\mu\nu} + \Lambda g_{\mu\nu} = \hat{\alpha}H_{\mu\nu},
\end{equation} 
where
\begin{equation}
H_{\mu\nu} = - 2\left(R R_{\mu\nu} - 2R_{\mu\alpha\nu\beta}R^{\alpha\beta} + R_{\mu\alpha\beta\sigma}
R_{\nu}^{ \ \alpha\beta\sigma} - 2R_{\mu\alpha}R_{\nu}^{ \ \alpha} - \frac{1}{4}g_{\mu\nu}\mathcal{G}\right).
\end{equation}
The right-hand side of the above equation must vanish in four dimensions because it originates from an antisymmetrized tensor over five indices~\cite{fernandes20224d}. However, as shown in Ref.~\cite{glavan2020einstein}, we have:
\begin{equation}
\label{eq:GB4Ddyna}
g^{\mu\nu}H_{\mu\nu} = \frac{1}{2}(D - 4)\mathcal{G} \, .
\end{equation}
This indicates that the multiplicative factor $(D - 4)$ is precisely cancelled by the proposed re-scaling of $\hat{\alpha}$, leaving a non-vanishing contribution to the trace of the field equations as $D \rightarrow 4$. Therefore, this clearly demonstrates that although the first term in the action (\ref{loop corrected action}) is divergent, it results in a non-trivial dynamics of spacetime. This arises from the non-trivial contribution on the right-hand side of Einstein's equations, as derived from the modified Einstein's equations in Eq.~(\ref{modified Einstein's equations}).
As mentioned earlier, there have been many attempts in the literature to have a systematic procedure for making the 4-D GB term non-topological; our analysis shows that coupling the massless scalar field to gravity can naturally lead to such a non-topological term. Our analysis shows that the effect of matter at the one-loop level generates a term related to the non-topological GB gravity.
To our knowledge, such an analysis has not been shown in the literature.

Second, the term inside the second parenthesis in \eqref{loop corrected action} acts as a counter-term, which must be added to the tree-level action to cancel this diverging term. These counter terms are different compared to other procedures in the literature like Kaluza-Klein dimensional reduction and conformal scaling~\cite{fernandes20224d}. To understand this further, let us first consider the Kaluza-Klein dimensional reduction procedure: In this case, the GB term with $1/(D - 4)$ coefficient can be regularised by adding a suitable counterterm (see, for example, \cite{fernandes20224d}). In the Kaluza-Klein dimensional reduction procedure of the action (13) in \cite{fernandes20224d}, the diverging term is of the form $\frac{1}{16\pi G}\frac{\alpha}{D - 4}\mathcal{G}$ which can be seen from the equation (69) in \cite{fernandes20224d}. In other words, in the $D \rightarrow 4$ limit, one needs to subtract the bare GB term of the same form. 

However, as mentioned above, this is not unique. For instance, in the conformal scaling scheme, one may add the term $\frac{1}{16\pi G}\frac{\alpha}{D - 4}\sqrt{-\tilde{g}}\tilde{\mathcal{G}}$ in the Lagrangian density which leads to following finite terms in the regularised action:
\begin{equation}
S = \frac{1}{16\pi G}\int\sqrt{-g}d^{4}x \Big[R + \alpha\left(4G^{\mu\nu}\nabla_{\mu}\phi\nabla_{\nu}\phi - \phi\mathcal{G} + 4\Box\phi(\nabla\phi)^{2} + 2(\nabla\phi)^{4}\right)\Big],
\end{equation}
where $\phi$ is the conformal factor connecting two frames by $\tilde{g}_{\mu\nu} = e^{2\phi}g_{\mu\nu}$. Note that both procedures lead to the same finite terms. However, these procedures do not provide a quantum field theoretic understanding for considering $\frac{\alpha}{D - 4}\mathcal{G}$ term in the action. Our procedure leads naturally to the dynamical GB term in 4-D. However, it also leads to additional diverging terms, which we can cancel by adding a suitable counterterm to the tree-level action.
Despite the non-renormalizability of GR, physical observables can be made finite by appropriately incorporating an infinite number of counterterms. It is important to understand that being a non-renormalizable theory does not diminish its physical significance. For a more detailed discussion on this topic, see section (12.3) in Ref.~\cite{book:91318327}.

While the analysis is classical, it is possible to get some insight from quantum field theory~\cite{book:91318327,Barth:1983hb}. In quantum field theory, regularization and renormalization are essential methods employed to address the divergences that emerge during the computation of physical quantities. These divergences typically occur when evaluating loop diagrams in quantum field interactions, resulting in divergent integrals that render the theory mathematically ill-posed.

It is essential to highlight a fundamental aspect of GR. While GR is a non-renormalizable theory, it remains feasible to render physical observables finite by systematically incorporating an infinite series of counterterms. This methodology underscores the notion that the non-renormalizability of a theory does not diminish its physical significance or applicability. A theory can retain its effectiveness in describing physical phenomena within a particular regime, even in the absence of renormalizability. (For a detailed discussion of this topic, see Section 12.3 of reference \cite{book:91318327}.) Specifically, the second term  (\ref{loop corrected action}):
%While  (\ref{loop corrected action}) includes both the scaled Gauss-Bonnet term (the first line), as proposed by Glavan and Lin, and an additional term quadratic in curvature. 
%
\[
- \alpha\int d^{4}x\sqrt{-g}\Big[8R_{\mu\nu}R^{\mu\nu} + 4R^{2}\Big],
\] 
arises in the loop-corrected quantum effective action. While this term is finite, the dimensional regularization technique, where $D = 4 - \epsilon$, introduces a prefactor $1/\epsilon$ that renders the term divergent. To eliminate this divergence, we add a counterterm to the Einstein-Hilbert action of the form:
\[
S_{ct} = \frac{\alpha}{\epsilon}\int d^{4}x\sqrt{-g}\Big[8R_{\mu\nu}R^{\mu\nu} + 4R^{2}\Big].
\] 
This counterterm cancels the divergence introduced by the second term in Equation (\ref{loop corrected action}). However, it also generates additional divergences in the loop-corrected quantum effective action. To address these new divergences, further counterterms must be introduced.

This iterative process is characteristic of non-renormalizable theories, such as GR, where an infinite number of counterterms is required to make physical observables finite in a perturbative framework. On the other hand, the first term in equation (\ref{loop corrected action}) leads to a finite correction to the effective action and the modified equations of motion. Therefore, this term is retained in the quantum-corrected action. Although our discussion here focuses on Equation (\ref{loop corrected action}), the same reasoning applies to the equation (\ref{fermion loop correction}).

This approach aligns with the results of Barth and Christensen in \cite{Barth:1983hb} 
where they investigated the quantization of fourth-order gravity theories using the functional integral formalism. Their analysis provides essential insights into one-loop counterterms in higher-derivative gravity.  Specifically, Eq. (C23) in Ref.~\cite{Barth:1983hb} demonstrates that divergent counterterms in position space, computed using the point-splitting method, are less severe in quartic gravity theories compared to those arising at the one-loop level in the Einstein-Hilbert action, which corresponds to quadratic gravity. These findings are instrumental in understanding the structure of higher-derivative counterterms and their role in quantum corrections to gravity. This is currently under investigation. Therefore, following our approach of adding suitable counterterms up to any given order, one may obtain a finite effective action with the scaled Gauss-Bonnet gravity term proposed by Glavan and Lin which modified the field equations associated with the metric variation.
 
Third, while the analysis is done for a massless scalar field, the results can be straightforwardly extended for self-interacting potential terms. Note that the subleading term $(\epsilon^0)$ does not modify the GB term.
It is important here to point out that in $D$ dimensions the number of massless graviton degrees of freedom in GR is $D(D - 3)/2$. Therefore, the number of extra degrees of freedom in $D$-dimension compared to four-dimension is $\frac{D(D - 3)}{2} - 2 = (D - 4)(D + 1)/2$ which is proportional to $\epsilon$ for $D = 4 - \epsilon$ dimensional regularization. This clearly shows that the number of degrees of freedom in the dimensional regularization is not going to be affected in the leading order correction.

\noindent \section{Effect of interaction between Gravitons and Dirac fermions at 1-loop}
%\label{sec:1loop-Dirac}
%
The above result naturally leads to the following questions: How generic is the above result? What happens to the Fermions? In this section,  we consider a massless Dirac field minimally coupled to gravity to address this. The action for such a field is:
\begin{equation}
S = \frac{1}{2\kappa^{2}}\int d^{4}x \ \sqrt{-g}R - \int d^{4}x \sqrt{-g}\bar{\psi}i
e_{ \ a}^{\mu}\gamma^{a}\mathcal{D}_{\mu}\psi,
\end{equation}
where $\psi$ is a massless Dirac field, $e_{ \ a}^{\mu}$ are tetrads satisfying the 
relation $g_{\mu\nu}e_{ \ a}^{\mu}e_{ \ b}^{\nu} = \eta_{ab}$, $\gamma^{a}$s are the  Dirac gamma matrices satisfying the Clifford algebra $\{\gamma^{a},\gamma^{b}\} = - 2\eta^{ab}$, and $\mathcal{D}_{\mu} = \partial_{\mu} + \Gamma_{\mu}$ is the spin-covariant derivative. 

Like in the previous section, to keep things transparent, we consider the first-order perturbations ($h_{\mu\nu}$) about the Minkowski background: 
$g_{\mu\nu} = \eta_{\mu\nu} + \kappa h_{\mu\nu}$. The expansion of the Einstein-Hilbert action up to the second order in the metric perturbation is identical. 
Following the approach in the previous section, we may write the matter action
as:
\begin{equation}
S_{M}[\psi] = - \int d^{4}x \bar{\psi}i\gamma^{\mu}\partial_{\mu}\psi + \kappa\int d^{4}x
h_{\mu\nu}T^{\mu\nu},
\end{equation} 
where the energy-momentum tensor of massless Dirac field theory in Minkowski spacetime 
is given by
\begin{equation}
T^{\mu\nu} = \frac{i}{4}\bar{\psi}[\gamma^{\mu}\overrightarrow{\partial}^{\nu} + \gamma^{\nu}\overrightarrow{\partial}^{\mu} - \gamma^{\mu}\overleftarrow{\partial}^{\nu} - \gamma^{\nu}
\overleftarrow{\partial}^{\mu}]\psi.
\end{equation}
\begin{figure}
\includegraphics[height = 2cm, width = 5cm]{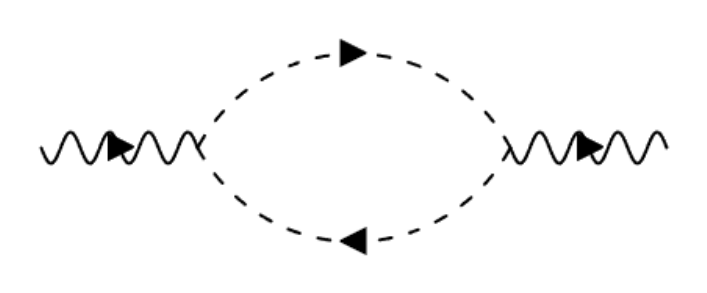}
\caption{One-loop Feynman diagram describing the self-energy correction to graviton propagator 
due to matter field}
\label{self-energy2}
\end{figure}

Like earlier, as described by the Feynman diagram in Fig. \ref{self-energy2}, we now compute the self-energy correction to graviton due to matter one-loop. 
In this case, the vertex in momentum space associated with the graviton-fermion interaction is given by:
\begin{equation}
\begin{split}
V^{\mu\nu}(p,q) & = - \frac{i\kappa}{8}[(p - q)^{\mu}\gamma^{\nu} + (p - q)^{\nu}\gamma^{\mu}
- \eta^{\mu\nu}(\slashed{p} - \slashed{q})]\\
 & \equiv - \frac{i\kappa}{8}\mathcal{P}_{ \ \ \rho\sigma}^{\mu\nu}(p - q)^{\rho}\gamma^{\sigma},
\end{split}
\end{equation}
where
\begin{equation}
\mathcal{P}_{ \ \ \rho\sigma}^{\mu\nu} = \delta_{\rho}^{\mu}\delta_{\sigma}^{\nu} + \delta
_{\sigma}^{\mu}\delta_{\rho}^{\nu} - \eta^{\mu\nu}\eta_{\rho\sigma}.
\end{equation}

To evaluate the self-energy corrections to gravitons, we evaluate the following integral corresponding to the above Feynman diagram \eqref{self-energy2}:
\begin{eqnarray}
& & \mathcal{J}^{\mu\nu\alpha\beta}  =  
- \mathcal{P}_{ \ \ \rho\sigma}^{\mu\nu}\mathcal{P}_{ \ \ 
\gamma\delta}^{\alpha\beta}\frac{\kappa^{2}}{64}\int\frac{d^{D}k}{(2\pi)^{D}}\text{Tr}\Big[(p + 2k)^{\rho}\gamma^{\sigma}\frac{1}{\slashed{p} + \slashed{k}} (p + 2k)^{\gamma}\gamma^{\delta}\frac{1}{\slashed{k}}\Big]
 \nonumber  \\
%& = - \mathcal{P}_{ \ \ \rho\sigma}^{\mu\nu}\mathcal{P}_{ \ \ \gamma\delta}^{\alpha\beta} \frac{\kappa^{2}}{64}\int\frac{d^{D}k}{(2\pi)^{D}}\frac{1}{k^{2}(p + k)^{2}}\text{Tr}[(p + 2k)^{\rho}\gamma^{\sigma}\\
% & \times (\slashed{p} + \slashed{k})(p + 2k)^{\gamma}\gamma^{\delta}\slashed{k}]\\
% & = - \mathcal{P}_{ \ \ \rho\sigma}^{\mu\nu}\mathcal{P}_{ \ \ \gamma\delta}^{\alpha\beta}
%\frac{\kappa^{2}}{64}\int\frac{d^{D}k}{(2\pi)^{D}}\frac{(p + 2k)^{\rho}(p + 2k)^{\gamma}} {k^{2}(p + k)^{2}}(p + k)_{\alpha}k_{\beta}\\
% & \times\text{Tr}[\gamma^{\sigma}\gamma^{\alpha}\gamma^{\delta}\gamma^{\beta}]\\
&& = - \mathcal{P}_{ \ \ \rho\sigma}^{\mu\nu}\mathcal{P}_{ \ \ \gamma\delta}^{\alpha\beta}
\frac{\kappa^{2}}{16}\int\frac{d^{D}k}{(2\pi)^{D}}\frac{(p + 2k)^{\rho}(p + 2k)^{\gamma}}
{k^{2}(p + k)^{2}}\nonumber  \\
\label{eq:1loppintegrals-Dirac}
 & & \qquad (p + k)_{\alpha}k_{\beta}  \times (\eta^{\sigma\alpha}\eta^{\delta\beta} - \eta^{\sigma\delta}\eta^{\alpha\beta}
 + \eta^{\sigma\beta}\eta^{\alpha\delta}). 
\end{eqnarray}
As shown in the Appendix \ref{1-Loop integrals}, we can explicitly evaluate up to $1/\epsilon$ order leading to:
\begin{equation}
- i h_{\mu\nu}(-p)\mathcal{J}^{\mu\nu\alpha\beta}h_{\alpha\beta}(p) = \frac{\kappa^{2} \beta}{4 \epsilon}  p^{4}[h_{\mu\nu}(- p)h^{\mu\nu}(p) - 2h(-p)h(p)] \, .
\end{equation}
Thus, the correction to the quantum effective action is:
\begin{equation}
S_{\text{corr}}^{(2)} = \frac{\beta}{4\epsilon}\int d^{4}x \Big[\kappa^{2}h_{\mu\nu}(\Box)^{2}
h^{\mu\nu} - 2\kappa^{2}h(\Box)^{2}h\Big] \, ,
\end{equation}
where 
\[
\beta = \frac{1}{5(4\pi)^{2}} \, .
\]

Like in the scalar field case, we can compare the above term with the 
quadratic gravity action \eqref{eq:GenQuadaction}.  Expanding the action 
\eqref{eq:GenQuadaction} to the second-order in the metric perturbations and comparing with the above correction action, we obtain the following relations
\begin{equation}
4a + b = \frac{\beta }{\epsilon}, \ c - a = - \frac{2\beta }{\epsilon} \, .
\end{equation}
Rewriting the coupling constants as
\begin{equation}
a = \frac{\beta }{\epsilon}\bar{a}, \ b = \frac{\beta }{\epsilon}\bar{b}, \ c = \frac{\beta }{\epsilon}\bar{c}, 
\end{equation}
we obtain the following relations
\begin{equation}
4\bar{a} + \bar{b} = 1, \ \bar{c} - \bar{a} = -2.
\end{equation}
Now, we can write the quantum correction to the effective action as follows
\begin{eqnarray}\label{fermion loop correction}
S_{corr}^{(2)} & =& \frac{\beta}{\epsilon} \bar{a}\int d^{4}x\sqrt{-g}\Big[R_{\mu\nu\rho\sigma}R^{\mu\nu\rho
\sigma} - 4R_{\mu\nu}R^{\mu\nu} + R^{2}\Big] \nonumber \\
 & +& \frac{\beta}{\epsilon} 
 \int d^{4}x\sqrt{-g}\Big[R_{\mu\nu}R^{\mu\nu} - 2R^{2}\Big], 
\end{eqnarray}
This is the second key result regarding which we want to discuss the following points: First, like in the scalar field case, the term inside the first integral in the RHS of the above expression is the GB term and is proportional to $(D-4)$~\cite{Padmanabhan:2013xyr}. Since the prefactor is proportional to $1/\epsilon = 1/(4 - D)$, $(D-4)$ factor that makes the GB term topological in 4-D cancels (cf. Eq.~\ref{eq:GB4Ddyna}), leading to a dynamical 4-D GB term. 
Our analysis shows that the minimal coupling of the matter field with gravity naturally leads to a non-topological term. Second, 
the term inside the second parenthesis acts as a counter-term, which must be added to the tree-level action to cancel this diverging term. It is important to note that $R_{\mu\nu}R^{\mu\nu}$ counter-term for the scalar and Dirac fields have opposite signs. {This is not unexpected since the one-loop computation of the Fermions involves trace operation with a negative sign due to the anti-commutation relation of the field variables. For this reason, the sign in front of $R_{\mu\nu}R^{\mu\nu}$ counter-term and the overall sign in front of the Gauss-Bonnet term is opposite compared to the scalar matter field.}

\noindent \section{Discussion}
GR fails to be universally applicable in strong-field gravity because the higher-derivative corrections to the Einstein-Hilbert action become significant in this regime. Consequently, various observables, such as the photon sphere around a black hole and the cataclysmic events like binary black hole mergers, can undergo significant alterations. The simplest form of a higher-order modified gravity theory is the Einstein-Gauss-Bonnet gravity theory, which possesses a second-order equation of motion and is devoid of ghosts. However, this theory is primarily topological in four dimensions, as previously mentioned, and classically does not modify the equation of motion. Conversely, Glavan and Lin proposed a mathematical prescription to render the non-topological Gauss-Bonnet gravity theory in four dimensions~\cite{glavan2020einstein}.

Nevertheless, this prescription appears ad-hoc at first glance and lacks a solid mathematical motivation from the quantum field theory perspective~\cite{fernandes20224d}. In this study, we have provided a well-defined mathematical approach to derive the above prescription. To obtain such a prescription, we have utilized the technique of dimensional regularization, commonly employed in quantum field theory, to regulate divergences. Our analysis commenced with the linearized Einstein-Hilbert action coupled to massless matter field theories, specifically scalar and fermionic matter. We computed the self-energy correction to gravitons due to matter up to one-loop. After incorporating appropriate counterterms, we obtained the Einstein-Gauss-Bonnet gravity theory with a dimensional regularization parameter. This result unequivocally demonstrates that the simplest version of Lovelock gravity, namely the Einstein-Gauss-Bonnet gravity theory, can be rendered non-topological in 4-D when accounting for the quantum effects of matter on gravity. Despite the non-renormalizability of GR, physical observables can be made finite by appropriately incorporating an infinite number of counterterms. It is important to understand that being a non-renormalizable theory does not diminish its physical significance. For a more detailed discussion on this topic, see section (12.3) in Ref.~\cite{book:91318327}.

The significance of our present work lies in its ability to introduce a novel perspective on the 4-D Einstein-Gauss-Bonnet gravity theory. To begin with, in the context of early Universe cosmology, it was shown that Higgs field non-minimally coupled to the Gauss-Bonnet term can lead to inflation with an exit depending on the energy scale~\cite{Mathew:2016anx}. With the non-topological Einstein-Gauss-Bonnet gravity theory now available in four dimensions, it is worth exploring whether a qualitatively similar conclusion can be reached without the presence of the Higgs field. In addition, the non-topological Gauss-Bonnet term is significant in investigating the formation and evaporation of primordial black holes in the early universe. This is because the Hawking temperature and Kretschmann scalar of an asymptotically flat space-time exhibit an inverse proportionality to the black hole mass. As a result, the Gauss-Bonnet term assumes a dominant role for primordial black holes, especially considering that their masses are anticipated to range between $10^{15}$ and $10^{26}$ grams \cite{kawai2021primordial, zhang2022primordial}.

Furthermore, our approach enables the examination of higher loop self-energy correction terms to ascertain whether higher-order Lovelock terms can be derived similarly~\cite{Padmanabhan:2013xyr}. 
Recently, it was argued that the three-point function can lead to GB term~\cite{Duff:2020dqb}. It will be interesting to repeat the analysis and see whether the higher order correlations can provide bounds on the GB coupling constant.
Also, it would be appealing to conduct a similar analysis while considering metric perturbations around a curved space-time in a covariant manner, especially for asymptotic non-flat space-times. 
In conclusion, it would be interesting if our work can be extended to the general effective field theory of gravity coupled to the Standard Model of particle physics \cite{Ruhdorfer:2019qmk}. This is currently under investigation.

The primary motivation of the work is to understand the origin of the GB term with the $1/(D - 4)$ coefficient, which is non-topological in the $D \rightarrow 4$ limit and not the strong coupling problem~\cite{fernandes20224d, Lu:2020iav, Kobayashi:2020wqy, Fernandes:2020nbq, Hennigar:2020lsl, Fernandes:2021dsb, Aoki:2020lig, Clifton:2020xhc, Bonifacio:2020vbk}. Our work explicitly demonstrates this using the dimensional regularisation procedure often used in quantum field theory. Interestingly, 
as shown in the Appendix \ref{Non-minimal term}, the non-minimal kinetic energy term mentioned can also be generated in the one-loop quantum correction due to the matter field. However, such a term in the quantum-corrected effective action does not give rise to a strong coupling as the tree-level action contains a canonical kinetic energy term of the matter field. As a result, our approach does not lead to strong coupling. We are currently investigating the effect of 2-loops on the strong coupling problem.

Lastly, we want to emphasize the mathematical consistency of our approach in deriving non-topological Gauss-Bonnet term in four dimensions through the dimensional regularization prescription. As part of our analysis, we have completely neglected the 3-point, 4-point, and other graviton vertices while computing the self-energy correction of gravitons from the expansion of the Einstein-Hilbert action. We want to stress that the self-energy term only contributes to the kinetic part of the effective action. Conversely, the self-energy corrections arising from the higher-order vertices mentioned above are of higher order in $\kappa^{2}$. Hence, those terms will contain higher derivatives as the overall effective action should be dimensionless. Therefore, our approach is mathematically consistent in deriving the non-topological Gauss-Bonnet term through dimensional regularization at the one-loop level within the linearized theory of gravity up to the second order in the metric function $h_{\mu\nu}$. 

\emph{Acknowledgement:-}
The authors thank S. Chakraborty and K. Lochan for discussions. SM is supported by SERB-Core Research Grant (Project: CRG/2022/002348).

%\bibliographystyle{apsrev}
%\bibliography{bibtexfile}

\appendix

\section{Linearised gravity}\label{App:Linearised gravity}
Although this is a standard calculation, for ease of verification, 
we compute the relevant quantities to obtain the linearised theory of gravity within the framework of Einstein's general relativity in this supplemental material.
In order to do that we consider the following decomposition of the metric
\begin{equation}\label{metric decomposition 1}
g_{\mu\nu} = \bar{g}_{\mu\nu} + \varepsilon h_{\mu\nu},
\end{equation}
where $\bar{g}_{\mu\nu}$ is the metric describing the background geometry \textit{w.r.t}
which we linearised the theory of gravity, $h_{\mu\nu}$ is the metric perturbation and
$\varepsilon$ carries the order of the metric perturbation. This leads to the following
expression of the inverse metric
\begin{equation}\label{inverse metric 1}
g^{\mu\nu} = \bar{g}^{\mu\nu} - \varepsilon h^{\mu\nu} + \varepsilon^{2}h^{\mu\rho}
h_{ \ \rho}^{\nu} + \mathcal{O}(\varepsilon^{3}),
\end{equation}
where $\bar{g}^{\mu\nu}$ is the inverse metric of the background spacetime, and $h^{\mu\nu}$
is a rank-2 contravariant tensor defined by contracting the metric perturbation \textit{w.r.t}
the inverse metric $\bar{g}^{\mu\nu}$. Given the above set of relation, we are now able to
compute the Christoffel symbol defined in the following manner
\begin{equation}\label{Christoffel symbol 1}
\Gamma_{ \ \mu\nu}^{\rho} = \frac{1}{2}g^{\rho\lambda}(\partial_{\mu}g_{\lambda\nu} + 
\partial_{\nu}g_{\mu\lambda} - \partial_{\lambda}g_{\mu\nu}),
\end{equation} 
which reduces to the following form after using the relations \eqref{metric decomposition 1}
and \eqref{inverse metric 1}
\begin{equation}
\begin{split}
\Gamma_{ \ \mu\nu}^{\rho} & = \bar{\Gamma}_{ \ \mu\nu}^{\rho} + \varepsilon\Big[\frac{1}{2}
\bar{g}^{\rho\lambda}(\partial_{\mu}h_{\lambda\nu} + \partial_{\nu}h_{\mu\lambda} - 
\partial_{\lambda}h_{\mu\nu})\\
 & - \bar{g}^{\rho\alpha}h_{\alpha\beta}\bar{\Gamma}_{ \ \mu\nu}^{\beta}\Big] - 
 \varepsilon^{2}\Big[\frac{1}{2}\bar{g}^{\beta\lambda}h_{ \ \beta}^{\rho}(\partial_{\mu}
 h_{\lambda\nu} + \partial_{\nu}h_{\mu\lambda} - \partial_{\lambda}h_{\mu\nu})\\
 & - \bar{g}^{\alpha\beta}h_{ \ \beta}^{\rho}h_{\alpha\sigma}\bar{\Gamma}_{ \ \mu\nu}^{\sigma}
 \Big], 
\end{split}
\end{equation}
where
\begin{equation}
\bar{\Gamma}_{ \ \mu\nu}^{\rho} = \frac{1}{2}\bar{g}^{\rho\lambda}(\partial_{\mu}\bar{g}
_{\lambda\nu} + \partial_{\nu}\bar{g}_{\mu\lambda} - \partial_{\lambda}\bar{g}_{\mu\nu}).
\end{equation}
After a change in the indices of the following form
\begin{equation}
\begin{split}
\bar{g}^{\rho\alpha}h_{\alpha\beta}\bar{\Gamma}_{ \ \mu\nu}^{\beta} & \rightarrow 
\bar{g}^{\rho\lambda}h_{\lambda\beta}\bar{\Gamma}_{ \ \mu\nu}^{\beta}\\
\bar{g}^{\alpha\beta}h_{ \ \beta}^{\rho}h_{\alpha\sigma}\bar{\Gamma}_{ \ \mu\nu}^{\sigma}
& \rightarrow \bar{g}^{\lambda\beta}h_{ \ \beta}^{\rho}h_{\lambda\sigma}\bar{\Gamma}_{ \ 
\mu\nu}^{\sigma},
\end{split}
\end{equation}
we obtain the following expression of Christoffel symbols
\begin{equation}
\begin{split}
\Gamma_{ \ \mu\nu}^{\rho} & = \bar{\Gamma}_{ \ \mu\nu}^{\rho} + \frac{\varepsilon}{2}
\bar{g}^{\rho\lambda}[\partial_{\mu}h_{\lambda\nu} + \partial_{\nu}h_{\mu\lambda} - 
\partial_{\lambda}h_{\mu\nu} - 2h_{\lambda\beta}\bar{\Gamma}_{ \ \mu\nu}^{\beta}]\\
 & - \frac{\varepsilon^{2}}{2}\bar{g}^{\beta\lambda}h_{ \ \beta}^{\rho}[\partial_{\mu}
 h_{\lambda\nu} + \partial_{\nu}h_{\mu\lambda} - \partial_{\lambda}h_{\mu\nu} - 
 2h_{\lambda\sigma}\bar{\Gamma}_{ \ \mu\nu}^{\sigma}].
\end{split}
\end{equation}
Now adding and subtracting the following terms at different order
\begin{equation}
\begin{split}
\mathcal{O}(\varepsilon): & h_{\mu\beta}\bar{\Gamma}_{ \ \nu\lambda}^{\beta} + h_{\nu\beta}
\bar{\Gamma}_{ \ \mu\lambda}^{\beta}\\
\mathcal{O}(\varepsilon^{2}): & h_{\mu\sigma}\bar{\Gamma}_{ \ \nu\lambda}^{\sigma} + 
h_{\nu\sigma}\bar{\Gamma}_{ \ \mu\lambda}^{\sigma}, 
\end{split}
\end{equation}
and then regrouping the terms gives
\begin{equation}
\begin{split}
\Gamma_{ \ \mu\nu}^{\rho} & = \bar{\Gamma}_{ \ \mu\nu}^{\rho} + \frac{\varepsilon}{2}
\bar{g}^{\rho\lambda}(\bar{\nabla}_{\mu}h_{\lambda\nu} + \bar{\nabla}_{\nu}h_{\mu\lambda}
- \bar{\nabla}_{\lambda}h_{\mu\nu})\\
 & - \frac{\varepsilon^{2}}{2}h_{ \ \beta}^{\rho}\bar{g}^{\beta\lambda}(\bar{\nabla}_{\mu}
 h_{\lambda\nu} + \bar{\nabla}_{\nu}h_{\mu\lambda} - \bar{\nabla}_{\lambda}h_{\mu\nu}),
\end{split}
\end{equation}
where we recognize the following
\begin{equation}
\bar{\nabla}_{\lambda}h_{\mu\nu} = \partial_{\lambda}h_{\mu\nu} - h_{\nu\sigma}\bar{\Gamma}
_{ \ \mu\lambda}^{\sigma} - h_{\mu\sigma}\bar{\Gamma}_{ \ \nu\lambda}^{\sigma}.
\end{equation}
Therefore, defining the linearised Christoffel symbol as
\begin{equation}
(\Gamma_{ \ \mu\nu}^{\rho})_{L} = \frac{1}{2}\bar{g}^{\rho\lambda}(\bar{\nabla}_{\mu}
h_{\lambda\nu} + \bar{\nabla}_{\nu}h_{\mu\lambda} - \bar{\nabla}_{\lambda}h_{\mu\nu}),
\end{equation}
we may now write
\begin{equation}
\Gamma_{ \ \mu\nu}^{\rho} = \bar{\Gamma}_{ \ \mu\nu}^{\rho} + \varepsilon(\Gamma_{ \ \mu\nu}
^{\rho})_{L} - \varepsilon^{2}h_{ \ \beta}^{\rho}(\Gamma_{ \ \mu\nu}^{\beta})_{L}.
\end{equation}
The Riemann curvature tensor is defined as
\begin{equation}
R_{ \ \nu\rho\sigma}^{\mu} = \partial_{\rho}\Gamma_{ \ \sigma\nu}^{\mu} + \Gamma_{ \ \rho
\lambda}^{\mu}\Gamma_{ \ \sigma\nu}^{\lambda} - \partial_{\sigma}\Gamma_{ \ \rho\nu}^{\mu}
- \Gamma_{ \ \sigma\lambda}^{\mu}\Gamma_{ \ \rho\nu}^{\lambda}.
\end{equation}
We start by defining the following quantity
\begin{equation}\label{defintion 1}
\delta\Gamma_{ \ \mu\nu}^{\rho} \equiv \Gamma_{ \ \mu\nu}^{\rho} - \bar{\Gamma}_{ \ \mu\nu}
^{\rho} = \varepsilon(\Gamma_{ \ \mu\nu}^{\rho})_{L} - \varepsilon^{2}h_{ \ \beta}^{\rho}
(\Gamma_{ \ \mu\nu}^{\beta})_{L},
\end{equation}
in terms of which we may express the Riemann curvature tensor as
\begin{equation}\label{Riemann tensor decomposition 1}
\begin{split}
R_{ \ \nu\rho\sigma}^{\mu} & = \bar{R}_{ \ \nu\rho\sigma}^{\mu} + (\partial_{\rho}\delta
\Gamma_{ \ \sigma\nu}^{\mu} + \bar{\Gamma}_{ \ \rho\lambda}^{\mu}\delta\Gamma_{ \ \sigma
\nu}^{\lambda} - \bar{\Gamma}_{ \ \rho\nu}^{\lambda}\delta\Gamma_{ \ \sigma\lambda}^{\mu})\\
 & - (\partial_{\sigma}\delta\Gamma_{ \ \rho\nu}^{\mu} + \bar{\Gamma}_{ \ \sigma\lambda}
 ^{\mu}\delta\Gamma_{ \ \rho\nu}^{\lambda} - \bar{\Gamma}_{ \ \sigma\nu}^{\lambda}\delta
 \Gamma_{ \ \rho\lambda}^{\mu})\\
 & + \delta\Gamma_{ \ \rho\lambda}^{\mu}\delta\Gamma_{ \ \sigma\nu}^{\lambda} - \delta\Gamma
_{ \ \sigma\lambda}^{\mu}\delta\Gamma_{ \ \rho\nu}^{\lambda},
\end{split}
\end{equation}
where
\begin{equation}
\bar{R}_{ \ \nu\rho\sigma}^{\mu} = \partial_{\rho}\bar{\Gamma}_{ \ \sigma\nu}^{\mu} + 
\bar{\Gamma}_{ \ \rho\lambda}^{\mu}\bar{\Gamma}_{ \ \sigma\nu}^{\lambda} - \partial_{\sigma}
\bar{\Gamma}_{ \ \rho\nu}^{\mu} - \bar{\Gamma}_{ \ \sigma\lambda}^{\mu}\bar{\Gamma}_{ \ \rho
\nu}^{\lambda}.
\end{equation}
Now adding and subtracting the term $\bar{\Gamma}_{ \ \rho\sigma}^{\lambda}\delta\Gamma_{ \ 
\lambda\nu}^{\mu}$ from the equation \eqref{Riemann tensor decomposition 1}, we obtain the 
following relation
\begin{equation}
R_{ \ \nu\rho\sigma}^{\mu} = \bar{R}_{ \ \nu\rho\sigma}^{\mu} + \bar{\nabla}_{\rho}\delta
\Gamma_{ \ \sigma\nu}^{\mu} - \bar{\nabla}_{\sigma}\delta\Gamma_{ \ \rho\nu}^{\mu} + \delta
\Gamma_{ \ \rho\lambda}^{\mu}\delta\Gamma_{ \ \sigma\nu}^{\lambda} - \delta\Gamma_{ \ \sigma
\lambda}^{\mu}\delta\Gamma_{ \ \rho\nu}^{\lambda}. 
\end{equation}
Now using the relation \eqref{defintion 1}, we now express the above relation as follows
\begin{equation}
\begin{split}
R_{ \ \nu\rho\sigma}^{\mu} & = \bar{R}_{ \ \nu\rho\sigma}^{\mu} + \varepsilon[\bar{\nabla}
_{\rho}(\Gamma_{ \ \sigma\nu}^{\mu})_{L} - \bar{\nabla}_{\sigma}(\bar{\Gamma}_{ \ \rho\nu}
^{\mu})_{L}]\\
 & - \varepsilon^{2}\{\bar{\nabla}_{\rho}[h_{ \ \beta}^{\mu}(\Gamma_{ \ \sigma\nu}^{\beta})
_{L}] - \bar{\nabla}_{\sigma}[h_{ \ \beta}^{\mu}(\Gamma_{ \ \rho\nu}^{\beta})_{L}]\\
 & - (\Gamma_{ \ \rho\lambda}^{\mu})_{L}(\Gamma_{ \ \sigma\nu}^{\lambda})_{L} + (\Gamma_{ \ 
 \sigma\lambda}^{\mu})_{L}(\Gamma_{ \ \rho\nu}^{\lambda})_{L}\} + \mathcal{O}(\varepsilon^{3}).
\end{split}
\end{equation}
The above expression can further be reduced in the following form
\begin{equation}
\begin{split}
R_{ \ \nu\rho\sigma}^{\mu} & = \bar{R}_{ \ \nu\rho\sigma}^{\mu} + \varepsilon\{\bar{\nabla}
_{\rho}(\Gamma_{ \ \sigma\nu}^{\mu})_{L} - \bar{\nabla}_{\sigma}(\Gamma_{ \ \rho\nu}^{\mu})
_{L}\}\\
 & - \varepsilon^{2}\{h_{ \ \beta}^{\mu}[\bar{\nabla}_{\rho}(\Gamma_{ \ \sigma\nu}^{\beta})
_{L} - \bar{\nabla}_{\sigma}(\Gamma_{ \ \rho\nu}^{\beta})_{L}]\\
 & + (\Gamma_{ \ \sigma\nu}^{\beta})_{L}\bar{\nabla}_{\rho}h_{ \ \beta}^{\mu} - (\Gamma_{ \ 
\rho\nu}^{\beta})_{L}\bar{\nabla}_{\sigma}h_{ \ \beta}^{\mu}\\
 & - (\Gamma_{ \ \rho\beta}^{\mu})_{L}(\Gamma_{ \ \sigma\nu}^{\beta})_{L} + (\Gamma_{ \ 
\sigma\beta}^{\mu})_{L}(\Gamma_{ \ \rho\nu}^{\beta})_{L}\}.
\end{split}
\end{equation}
Defining the following quantity
\begin{equation}
(R_{ \ \nu\rho\sigma}^{\mu})_{L} = \bar{\nabla}_{\rho}(\Gamma_{ \ \sigma\nu}^{\mu})_{L} - 
\bar{\nabla}_{\sigma}(\Gamma_{ \ \rho\nu}^{\mu})_{L},
\end{equation}
we further reduce the expression of Riemann curvature tensor as
\begin{equation}
\begin{split}
R_{ \ \nu\rho\sigma}^{\mu} & = \bar{R}_{ \ \nu\rho\sigma}^{\mu} + \varepsilon(R_{ \ \nu\rho
\sigma}^{\mu})_{L} - \varepsilon^{2}\{h_{ \ \beta}^{\mu}(R_{ \ \nu\rho\sigma}^{\beta})_{L}\\
 & + (\Gamma_{ \ \sigma\nu}^{\beta})_{L}[\bar{\nabla}_{\rho}h_{ \ \beta}^{\mu} - (\Gamma_{ \ 
 \rho\beta}^{\mu})_{L}]\\
 & - (\Gamma_{ \ \rho\nu}^{\beta})_{L}[\bar{\nabla}_{\sigma}h_{ \ \beta}^{\mu} - (\Gamma_{ \ 
\sigma\beta}^{\mu})_{L}]\}.
\end{split}
\end{equation}
The term $* \equiv (\Gamma_{ \ \sigma\nu}^{\beta})_{L}[\bar{\nabla}_{\rho}h_{ \ \beta}^{\mu} 
- (\Gamma_{ \ \rho\beta}^{\mu})_{L}]$ can be expressed as
\begin{equation}
\begin{split}
* & = (\Gamma_{ \ \sigma\nu}^{\beta})_{L}\Big[\bar{\nabla}_{\rho}(\bar{g}^{\mu\alpha}h_{\alpha
\beta}) - \frac{1}{2}\bar{g}^{\mu\alpha}(\bar{\nabla}_{\rho}h_{\alpha\beta} + \bar{\nabla}_{\beta}
h_{\rho\alpha} - \bar{\nabla}_{\alpha}h_{\rho\beta})\Big]\\
 & = (\Gamma_{ \ \sigma\nu}^{\beta})_{L}\Big[\frac{1}{2}\bar{g}^{\mu\alpha}(\bar{\nabla}_{\rho}
h_{\alpha\beta} + \bar{\nabla}_{\alpha}h_{\rho\beta} - \bar{\nabla}_{\beta}h_{\rho\alpha})\Big]\\
 & = (\Gamma_{ \ \sigma\nu}^{\beta})_{L}\Big[\frac{1}{2}\bar{g}^{\mu\alpha}\delta_{\beta}^{ \ 
\lambda}(\bar{\nabla}_{\rho}h_{\alpha\lambda} + \bar{\nabla}_{\alpha}h_{\rho\lambda} - \bar{\nabla}
_{\lambda}h_{\rho\alpha})\Big]\\
 & = (\Gamma_{ \ \sigma\nu}^{\beta})_{L}\Big[\frac{1}{2}\bar{g}^{\mu\alpha}\bar{g}_{\beta\gamma}
 \bar{g}^{\gamma\lambda}(\bar{\nabla}_{\rho}h_{\alpha\lambda} + \bar{\nabla}_{\alpha}h_{\rho\lambda} 
 - \bar{\nabla}_{\lambda}h_{\rho\alpha})\Big]\\
 & = \bar{g}^{\mu\alpha}\bar{g}_{\beta\gamma}(\Gamma_{ \ \sigma\nu}^{\beta})_{L}(\Gamma_{ \ \rho
 \alpha}^{\gamma})_{L}.
\end{split}
\end{equation}
In the similar manner, we may express the following
\begin{equation}
(\Gamma_{ \ \rho\nu}^{\beta})_{L}[\bar{\nabla}_{\sigma}h_{ \ \beta}^{\mu} - (\Gamma_{ \ \sigma
\beta}^{\mu})_{L}] = \bar{g}^{\mu\alpha}\bar{g}_{\beta\gamma}(\Gamma_{ \ \rho\nu}^{\beta})_{L}
(\Gamma_{ \ \sigma\alpha}^{\gamma})_{L}.
\end{equation}
Using the above results, the final form of the Riemann tensor perturbed up to second order in 
metric perturbation is given as
\begin{equation}
\begin{split}
R_{ \ \nu\rho\sigma}^{\mu} & = \bar{R}_{ \ \nu\rho\sigma}^{\mu} + \varepsilon(R_{ \ \nu\rho\sigma}^{\mu})_{L} - \varepsilon^{2}\{h_{ \ \beta}^{\mu}(R_{ \ \nu\rho\sigma}^{\beta})_{L}\\
 & + \bar{g}^{\mu\alpha}\bar{g}_{\beta\gamma}[(\Gamma_{ \ \sigma\nu}^{\beta})_{L}(\Gamma_{ \ \rho
\alpha}^{\gamma})_{L} - (\Gamma_{ \ \rho\nu}^{\beta})_{L}(\Gamma_{ \ \sigma\alpha}^{\gamma})_{L}]\}.
\end{split}
\end{equation}
As a result, the Ricci tensor can be expressed as
\begin{equation}
\begin{split}
R_{\nu\sigma} & = \bar{R}_{\nu\sigma} + \varepsilon(R_{\nu\sigma})_{L} - \varepsilon^{2}
\{h_{ \ \beta}^{\mu}(R_{ \ \nu\mu\sigma}^{\beta})_{L}\\
 & + \bar{g}^{\mu\alpha}\bar{g}_{\beta\gamma}[(\Gamma_{ \ \sigma\nu}^{\beta})_{L}(\Gamma_{ \ \mu
\alpha}^{\gamma})_{L} - (\Gamma_{ \ \mu\nu}^{\beta})_{L}(\Gamma_{ \ \sigma\alpha}^{\gamma})_{L}]\}.
\end{split}
\end{equation}
From the definition of linearised Riemann tensor, we have the following expression
\begin{equation}
\begin{split}
(R_{ \ \nu\rho\sigma}^{\mu})_{L} & = \frac{1}{2}[\bar{\nabla}_{\rho}\bar{\nabla}_{\sigma}h_{ \ 
\nu}^{\mu} + \bar{\nabla}_{\rho}\bar{\nabla}_{\nu}h_{ \ \sigma}^{\mu} - \bar{\nabla}_{\rho}
\bar{\nabla}^{\mu}h_{\sigma\nu}\\
 & - \bar{\nabla}_{\sigma}\bar{\nabla}_{\rho}h_{ \ \nu}^{\mu} - \bar{\nabla}_{\sigma}\bar{\nabla}
_{\nu}h_{ \ \rho}^{\mu} + \bar{\nabla}_{\sigma}\bar{\nabla}^{\mu}h_{\rho\nu}],
\end{split}
\end{equation}
which leads to the following expression of $(R_{\nu\sigma})_{L}$
\begin{equation}
\begin{split}
(R_{\nu\sigma})_{L} & = (R_{ \ \nu\mu\sigma}^{\mu})_{L}\\
 & = \frac{1}{2}(\bar{\nabla}_{\mu}\bar{\nabla}_{\sigma}h_{ \ \nu}^{\mu} + \bar{\nabla}_{\mu}
 \bar{\nabla}_{\nu}h_{ \ \sigma}^{\mu} - \bar{\Box}h_{\sigma\nu} - \bar{\nabla}_{\sigma}\bar{
 \nabla}_{\nu}h).
\end{split}
\end{equation}
Therefore, the Ricci scalar can now be expressed as
\begin{equation}
\begin{split}
R & = (\bar{g}^{\nu\sigma} - \varepsilon h^{\nu\sigma} + \varepsilon^{2}h^{\nu\lambda}h_{ \ 
\lambda}^{\sigma})[\bar{R}_{\nu\sigma} + \varepsilon(R_{\nu\sigma})_{L}\\
 & - \varepsilon^{2}\{h_{ \ \beta}^{\mu}(R_{ \ \nu\mu\sigma}^{\beta})_{L} + \bar{g}^{\mu\alpha}
 \bar{g}_{\beta\gamma}[(\Gamma_{ \ \sigma\nu}^{\beta})_{L}(\Gamma_{ \ \mu\alpha}^{\gamma})_{L}\\
 & - (\Gamma_{ \ \mu\nu}^{\beta})_{L}(\Gamma_{ \ \sigma\alpha}^{\gamma})_{L}]\}]\\
 & = \bar{g}^{\nu\sigma}\bar{R}_{\nu\sigma} + \varepsilon[\bar{g}^{\nu\sigma}(R_{\nu\sigma})_{L}
 - h^{\nu\sigma}\bar{R}_{\nu\sigma}]\\
 & - \varepsilon^{2}[\bar{g}^{\nu\sigma}\{h_{ \ \beta}^{\mu}(R_{ \ \nu\mu\sigma}^{\beta})_{L} + 
 \bar{g}^{\mu\alpha}\bar{g}_{\beta\gamma}[(\Gamma_{ \ \sigma\nu}^{\beta})_{L}(\Gamma_{ \ \mu\alpha}
^{\gamma})_{L}\\
 & - (\Gamma_{ \ \mu\nu}^{\beta})_{L}(\Gamma_{ \ \sigma\alpha}^{\gamma})_{L}]\} + h^{\nu\sigma}
 (R_{\nu\sigma})_{L} - h^{\nu\lambda}h_{ \ \lambda}^{\sigma}\bar{R}_{\nu\sigma}] + \mathcal{O}
(\varepsilon^{3}).
\end{split}
\end{equation}
Using the following definitions
\begin{equation}
\bar{R} = \bar{g}^{\nu\sigma}\bar{R}_{\nu\sigma}, \ R_{L} = \bar{g}^{\nu\sigma}(R_{\nu\sigma})_{L}
- h^{\nu\sigma}\bar{R}_{\nu\sigma},
\end{equation}
and neglecting the higher order contributions, we obtain the following form of Ricci scalar
\begin{equation}
\begin{split}
R & = \bar{R} + \varepsilon R_{L} - \varepsilon^{2}\{\bar{g}^{\nu\sigma}h_{ \ \beta}^{\mu}
(R_{ \ \nu\mu\sigma}^{\beta})_{L} + h^{\nu\sigma}(R_{\nu\sigma})_{L}\\
 & - h^{\nu\lambda}h_{ \ \lambda}^{\sigma}\bar{R}_{\nu\sigma} + \bar{g}^{\nu\sigma}\bar{g}
^{\mu\alpha}\bar{g}_{\beta\gamma}[(\Gamma_{ \ \sigma\nu}^{\beta})_{L}(\Gamma_{ \ \mu\alpha}
^{\gamma})_{L}\\
 & - (\Gamma_{ \ \mu\nu}^{\beta})_{L}(\Gamma_{ \ \sigma\alpha}^{\gamma})_{L}]\}.
\end{split}
\end{equation}

\section{Linearised gravity in Einstein-Hilbert action}
\label{App:Linearised gravity EH}
In this appendix, we describe the linearised theory of Einstein's general relativity by expressing the Einstein-Hilbert action in terms of metric perturbation \textit{w.r.t} the Minkowski flat spacetime. The Einstein-Hilbert action is expressed as
\begin{equation}\label{E-H action}
S_{EH} = \frac{1}{2\kappa^{2}}\int d^{4}x \sqrt{-g}R,
\end{equation}
where $g$ is the determinant of the metric of the spacetime and $\kappa^{2} = 8\pi G$ in natural 
units. In order to make the above-mentioned expansion, we consider the following decomposition 
of the metric tensor
\begin{equation}
g_{\mu\nu} = \eta_{\mu\nu} + \kappa h_{\mu\nu},
\end{equation}
which leads to the following inverse metric
\begin{equation}
g^{\mu\nu} = \eta^{\mu\nu} - \kappa h^{\mu\nu}.
\end{equation}
Moreover, we also obtain the following relation
\begin{equation}
\sqrt{-g} \approx 1 + \frac{\kappa}{2}h,
\end{equation}
where $h = \eta^{\mu\nu}h_{\mu\nu}$. Hence, the expression of the Christoffel symbols
become
\begin{equation}
\Gamma_{ \ \mu\nu}^{\rho} = \frac{\kappa}{2}(\eta^{\rho\gamma} - \kappa h^{\rho\gamma})
(\partial_{\nu}h_{\mu\gamma} + \partial_{\mu}h_{\nu\gamma} - \partial_{\gamma}h_{\mu\nu}).
\end{equation}
As a result, up to first order, the expression of Christoffel symbols reduce to the
following form
\begin{equation}
\Gamma_{ \ \ \ \mu\nu}^{(1)\rho} = \frac{\kappa}{2}(\partial_{\nu}h_{\mu}^{ \ \rho} + \partial
_{\mu}h_{\nu}^{ \ \rho} - \partial^{\rho}h_{\mu\nu}),
\end{equation}
which leads to the following two relations
\begin{equation}
\begin{split}
\partial_{\rho}\Gamma_{ \ \ \ \mu\nu}^{(1)\rho} & = \frac{\kappa}{2}(\partial_{\rho}\partial
_{\nu}h_{\mu}^{ \ \rho} + \partial_{\rho}\partial_{\mu}h_{\nu}^{ \ \rho} - \partial_{\rho}
\partial^{\rho}h_{\mu\nu})\\
\partial_{\nu}\Gamma_{ \ \ \ \mu\rho}^{(1)\rho} & = \frac{\kappa}{2}(\partial_{\nu}\partial
_{\rho}h_{ \ \mu}^{\rho} + \partial_{\nu}\partial_{\mu}h_{ \ \rho}^{\rho} - \partial_{\nu}
\partial^{\rho}h_{\mu\rho}).
\end{split}
\end{equation}
As a result, up to first order, the Ricci tensor can be expressed as
\begin{equation}
\begin{split}
R_{\mu\nu}^{(1)} & = \partial_{\rho}\Gamma_{ \ \ \ \mu\nu}^{(1)\rho} - \partial_{\nu}\Gamma
_{ \ \ \ \mu\rho}^{(1)\rho}\\
 & = \frac{\kappa}{2}(\partial_{\rho}\partial_{\mu}h_{\nu}^{ \ \rho} + \partial_{\nu}\partial
^{\rho}h_{\mu\rho} - \partial_{\nu}\partial_{\mu}h - \Box h_{\mu\nu}),
\end{split}
\end{equation}
where $\Box = \partial_{\mu}\partial^{\mu}$. In order to obtain the second-order term for the
Ricci tensor, we now use the following form of the second-order Christoffel symbols
\begin{equation}
\Gamma_{ \ \ \ \mu\nu}^{(2)\rho} = - \frac{\kappa^{2}}{2}h^{\rho\gamma}(\partial_{\nu}h_{\mu
\gamma} + \partial_{\mu}h_{\nu\gamma} - \partial_{\gamma}h_{\mu\nu}).
\end{equation}
Using the above expression, we obtain the following results
\begin{equation}
\begin{split}
\partial_{\rho}\Gamma_{ \ \ \ \mu\nu}^{(2)\rho} & = - \frac{\kappa^{2}}{2}[\partial_{\rho}
h^{\rho\gamma}(\partial_{\nu}h_{\mu\gamma} + \partial_{\mu}h_{\nu\gamma} - \partial_{\gamma}
h_{\mu\nu})\\
 & + h^{\rho\gamma}\partial_{\rho}(\partial_{\nu}h_{\mu\gamma} + \partial_{\mu}h_{\nu\gamma} 
 - \partial_{\gamma}h_{\mu\nu})]\\
\partial_{\nu}\Gamma_{ \ \ \ \mu\rho}^{(2)\rho} & = - \frac{\kappa^{2}}{2}[\partial_{\nu}
h^{\rho\gamma}\partial_{\mu}h_{\rho\gamma} + h^{\rho\gamma}\partial_{\nu}\partial_{\mu}
h_{\rho\gamma}],
\end{split}
\end{equation}
which eventually leads to the following result after some algebra
\begin{equation}
\begin{split}
\partial_{\rho}\Gamma_{ \ \ \ \mu\nu}^{(2)\rho} - \partial_{\nu}\Gamma_{ \ \ \ \mu\rho}^{(2)
\rho} & = - \frac{\kappa^{2}}{2}[\partial_{\rho}h^{\rho\gamma}(\partial_{\nu}h_{\mu\gamma} + 
\partial_{\mu}h_{\nu\gamma} - \partial_{\gamma}h_{\mu\nu})\\
 & + h^{\rho\gamma}(\partial_{\rho}\partial_{\nu}h_{\mu\gamma} + \partial_{\rho}\partial_{\mu}
 h_{\nu\gamma} - \partial_{\rho}\partial_{\gamma}h_{\mu\nu}\\
 & - \partial_{\nu}\partial_{\mu}h_{\rho\gamma}) - \partial_{\nu}h^{\rho\gamma}\partial_{\mu}
 h_{\rho\gamma}].
\end{split}
\end{equation}
Moreover, we also have the following relations
\begin{equation}
\begin{split}
\Gamma_{ \ \ \ \gamma\rho}^{(1)\rho}\Gamma_{ \ \ \ \mu\nu}^{(1)\gamma} & = \frac{\kappa^{2}}{4}
[\partial_{\gamma}h(\partial_{\mu}h_{\nu}^{ \ \gamma} + \partial_{\nu}h_{\mu}^{ \ \gamma} - 
\partial^{\gamma}h_{\mu\nu})]\\
\Gamma_{ \ \ \ \gamma\nu}^{(1)\rho}\Gamma_{ \ \ \ \mu\rho}^{(1)\gamma} & = \frac{\kappa^{2}}{2}
\Big[\partial_{\gamma}h_{\nu}^{ \ \rho}(\partial_{\rho}h_{\mu}^{ \ \gamma} - \partial^{\gamma}
h_{\rho\mu}) + \frac{1}{2}\partial_{\nu}h_{\gamma}^{ \ \rho}\partial_{\mu}h_{\rho}^{ \ \gamma}
\Big].
\end{split}
\end{equation}
Combining all these pieces gives us the total Ricci tensor in the second-order: 
\begin{equation}
\begin{split}
R_{\mu\nu}^{(2)} & = - \frac{\kappa^{2}}{2}\Big[\partial_{\rho}h^{\rho\gamma}(\partial_{\nu}
h_{\mu\gamma} + \partial_{\mu}h_{\nu\gamma} - \partial_{\gamma}h_{\mu\nu})\\
 & + h^{\rho\gamma}(\partial_{\rho}\partial_{\nu}h_{\mu\gamma} + \partial_{\rho}\partial_{\mu}
 h_{\nu\gamma} - \partial_{\rho}\partial_{\gamma}h_{\mu\nu} - \partial_{\nu}\partial_{\mu}
 h_{\rho\gamma})\\
 & - \frac{1}{2}\partial_{\nu}h^{\rho\gamma}\partial_{\mu}h_{\rho\gamma} - \frac{1}{2}\partial
_{\gamma}h(\partial_{\mu}h_{\nu}^{ \ \gamma} + \partial_{\nu}h_{\mu}^{ \ \gamma} - \partial
^{\gamma}h_{\mu\nu})\\
 & + \partial_{\gamma}h_{\nu}^{ \ \rho}(\partial_{\rho}h_{\mu}^{ \ \gamma} - \partial^{\gamma}
 h_{\rho\mu})\Big].
\end{split}
\end{equation}
As a result, the Einstein-Hilbert action reduces to the following form up to quadratic order
\begin{equation}
\begin{split}
S_{EH}^{(1)} + S_{EH}^{(2)} & = \frac{1}{2\kappa^{2}}\int d^{4}x \left(1 + \frac{\kappa}{2}h
\right)(\eta^{\mu\nu} - \kappa h^{\mu\nu})(R_{\mu\nu}^{(1)} + R_{\mu\nu}^{(2)})\\
 & = \frac{1}{2\kappa^{2}}\int d^{4}x \Big[\eta^{\mu\nu}R_{\mu\nu}^{(1)} - \kappa h^{\mu\nu}
 R_{\mu\nu}^{(1)} + \frac{\kappa}{2}h\eta^{\mu\nu}R_{\mu\nu}^{(1)}\\
 & + \eta^{\mu\nu}R_{\mu\nu}^{(2)}\Big],
\end{split}
\end{equation}
where
\begin{equation}
\begin{split}
\eta^{\mu\nu}R_{\mu\nu}^{(1)} & = \kappa(\partial_{\rho}\partial_{\mu}h^{\rho\mu} - \Box h)\\
- \kappa h^{\mu\nu}R_{\mu\nu}^{(1)} & = - \frac{\kappa^{2}}{2}h^{\mu\nu}(\partial_{\rho}
\partial_{\mu}h_{\nu}^{ \ \rho} + \partial_{\nu}\partial^{\rho}h_{\mu\rho} - \partial_{\nu}
\partial_{\mu}h - \Box h_{\mu\nu})\\
\frac{\kappa}{2}h\eta^{\mu\nu}R_{\mu\nu}^{(1)} & = \frac{\kappa^{2}}{2}(h\partial_{\rho}
\partial_{\mu}h^{\rho\mu} - h\Box h)\\
\eta^{\mu\nu}R_{\mu\nu}^{(2)} & = - \frac{\kappa^{2}}{2}h\partial^{\gamma}\partial^{\mu}
h_{\mu\gamma} + \frac{\kappa^{2}}{4}h\Box h + \frac{\kappa^{2}}{2}h^{\rho\gamma}\partial
_{\rho}\partial^{\mu}h_{\mu\gamma}\\
 & - \frac{\kappa^{2}}{4}h^{\rho\gamma}\Box h_{\rho\gamma}.
\end{split}
\end{equation}
Collecting all these contributions, neglecting boundary terms and scaling by a constant, 
the Einstein-Hilbert action up to second-order reduces to the following form
\begin{equation}
\begin{split}
S_{EH} & = \int d^{D}x \Big[ - \frac{1}{2}\partial_{\lambda}h_{\mu\nu}\partial^{\lambda}
h^{\mu\nu} + \partial_{\mu}h_{\nu\lambda}\partial^{\nu}h^{\mu\lambda} - \partial_{\mu}
h^{\mu\nu}\partial_{\nu}h\\
 &  + \frac{1}{2}\partial_{\lambda}h\partial^{\lambda}h\Big].
\end{split}
\end{equation}
We may note that the above action is invariant under the following gauge transformation 
\begin{equation}
h_{\mu\nu} \rightarrow h_{\mu\nu} + \partial_{\mu}\xi_{\nu} + \partial_{\nu}\xi_{\mu}.
\end{equation}
The above action can also be expressed as
\begin{equation}
\begin{split}
S_{EH} & = \int d^{D}x \frac{1}{2}h_{\mu\nu}\mathcal{E}^{\mu\nu,\alpha\beta}h_{\alpha\beta}\\
\mathcal{E}_{ \ \ \ \alpha\beta}^{\mu\nu,} & = \left(\delta_{\alpha}^{(\mu}\delta_{\beta}^{\nu)}
- \eta^{\mu\nu}\eta_{\alpha\beta}\right)\Box - 2\partial^{(\mu}\partial_{(\alpha}\delta_{\beta)}
^{\nu)} + \partial^{\mu}\partial^{\nu}\eta_{\alpha\beta}\\
 & + \partial_{\alpha}\partial_{\beta}\eta^{\mu\nu}. 
\end{split}
\end{equation}
We first note the action of the kinetic operator $\mathcal{E}^{\mu\nu,\alpha\beta}$ on a general
symmetric tensor $\mathcal{Z}_{\alpha\beta}$
\begin{equation}
\begin{split}
\mathcal{E}^{\mu\nu,\alpha\beta}\mathcal{Z}_{\alpha\beta} & = \Box\mathcal{Z}^{\mu\nu} - \eta
^{\mu\nu}\Box\mathcal{Z} - 2\partial^{(\mu}\partial_{\alpha}\mathcal{Z}^{\nu)\alpha}\\
 & + \partial^{\mu}\partial^{\nu}\mathcal{Z} + \eta^{\mu\nu}\partial_{\alpha}\partial_{\beta}
 \mathcal{Z}^{\alpha\beta}, 
\end{split}
\end{equation}
which is a transverse tensor since
\begin{equation}
\partial_{\mu}(\mathcal{E}^{\mu\nu,\alpha\beta}\mathcal{Z}_{\alpha\beta}) = 0,
\end{equation}
which can be checked quite easily. Furthermore, the kinetic operator would annihilate any pure 
gauge 
\begin{equation}
\begin{split}
\mathcal{E}^{\mu\nu,\alpha\beta} & (\partial_{\alpha}\xi_{\beta} + \partial_{\beta}\xi_{\alpha})
 = \Box(\partial^{\mu}\xi^{\nu} + \partial^{\nu}\xi^{\mu}) - 2\eta^{\mu\nu}\Box\partial_{\alpha}
 \xi^{\alpha}\\
 & - \partial^{\mu}\partial_{\alpha}(\partial^{\nu}\xi^{\alpha} + \partial^{\alpha}
 \xi^{\nu})- \partial^{\nu}\partial_{\alpha}(\partial^{\mu}\xi^{\alpha} + \partial^{\alpha}
 \xi^{\mu})\\
 & + 2\partial^{\mu}\partial^{\nu}\partial_{\alpha}\xi^{\alpha} + \eta^{\mu\nu}\partial_{\alpha}
 \partial_{\beta}(\partial^{\alpha}\xi^{\beta} + \partial^{\beta}\xi^{\alpha}) = 0.
\end{split}
\end{equation}
As a result, the determinant of the kinetic operator vanishes which means it cannot be invertible.
Therefore, in order to find the propagator in this theory, we must fix the gauge freedom in this
theory. Here we choose the Lorenz gauge which is given by the following relation
\begin{equation}
\partial^{\mu}h_{\mu\nu} - \frac{1}{2}\partial_{\nu}h = 0.
\end{equation}
Therefore, given a metric perturbation, one can always do a gauge transformation
\begin{equation}
\delta h_{\mu\nu} = \partial_{\mu}\xi_{\nu} + \partial_{\nu}\xi_{\mu},
\end{equation}
such that  
\begin{equation}
\Box\xi_{\mu} = - \left(\partial^{\nu}h_{\mu\nu} - \frac{1}{2}\partial_{\mu}h\right),
\end{equation}
which makes sure the gauge fixing condition is satisfied. However, this fixes the gauge 
up to gauge transformation $\xi_{\mu}$ satisfy the harmonicity condition $\Box\xi_{\mu} 
= 0$. Besides that with this gauge choice, the equation of motion reduces to the following 
form
\begin{equation}
\Box h_{\mu\nu} - \frac{1}{2}\eta_{\mu\nu}\Box h = 0,
\end{equation}
and these are the gauge-fixed equations of motion. In order to obtain these equations of
motion, we have to add a gauge fixing Lagrangian density which is of the following form
\begin{equation}
\mathcal{L}_{GF} = - \left(\partial^{\nu}h_{\mu\nu} - \frac{1}{2}\partial_{\mu}h\right)^{2}.
\end{equation}
By doing that we obtain a neg Lagrangian density of the following form
\begin{equation}
\mathcal{L}' = \frac{1}{2}h_{\mu\nu}\Box h^{\mu\nu} - \frac{1}{4}h\Box h = \frac{1}{2}
h_{\mu\nu}\mathcal{O}^{\mu\nu,\alpha\beta}h_{\alpha\beta},
\end{equation}
where the operator is
\begin{equation}
\mathcal{O}^{\mu\nu,\alpha\beta} = \Box\Big[\frac{1}{2}(\eta^{\mu\alpha}\eta^{\nu\beta}
+ \eta^{\mu\beta}\eta^{\nu\alpha}) - \frac{1}{2}\eta^{\mu\nu}\eta^{\alpha\beta}\Big].
\end{equation}
In momentum space, the above operator reduces to the inverse of the propagator which is
of the following form
\begin{equation}
\mathcal{O}^{\mu\nu,\alpha\beta}(p) = - p^{2}\Big[\frac{1}{2}(\eta^{\mu\alpha}\eta^{\nu\beta}
+ \eta^{\mu\beta}\eta^{\nu\alpha}) - \frac{1}{2}\eta^{\mu\nu}\eta^{\alpha\beta}\Big]. 
\end{equation}
The propagator $\mathcal{D}_{\alpha\beta,\sigma\lambda}$ satisfies the relation
\begin{equation}
\mathcal{O}^{\mu\nu,\alpha\beta}\mathcal{D}_{\alpha\beta,\sigma\lambda} = \frac{i}{2}\left(
\delta_{\sigma}^{\mu}\delta_{\lambda}^{\nu} + \delta_{\sigma}^{\nu}\delta_{\lambda}^{\mu}
\right).
\end{equation}
By solving the above equation, we obtain the massless graviton propagator
\begin{equation}
\mathcal{D}_{\alpha\beta,\sigma\lambda}(p) = - \frac{i}{p^{2}}\Big[\frac{1}{2}(\eta_{\alpha
\sigma}\eta_{\beta\lambda} + \eta_{\alpha\lambda}\eta_{\beta\sigma}) - \frac{1}{D - 2}\eta
_{\alpha\beta}\eta_{\sigma\lambda}\Big].
\end{equation}
\section{1-Loop vertex integrals for scalars and Dirac fields}\label{1-Loop integrals}

In this Appendix, we explicitly evaluate the integrals in Secs. (2) and (3) up to $1/\epsilon$ order where 
$\epsilon = 4 - D$.

\subsection{Scalar fields}
\label{App:1-LoopScalar}
Eq. (8) is the self-energy corrections to gravitons due to massless scalar fields. To the leading order in $1/\epsilon$, the five integrals are given below:
\begin{equation}
\begin{split}
I_{\rho\sigma\gamma\delta}^{(1)} & = \int_{0}^{1}dx\int\frac{d^{D}k}{(2\pi)^{D}}\frac{p_{\rho}
p_{\gamma}k_{\sigma}k_{\delta}(1 - 2x)^{2}}{[k^{2} + p^{2}x(1 - x)]^{2}}\\
 & = - \frac{i}{30(4\pi)^{2}\epsilon}p_{\rho}p_{\gamma}p^{2}\eta_{\sigma\delta} + \mathcal{O}
 (\epsilon^{0})\\
I_{\rho\sigma\gamma\delta}^{(2)} & = \int_{0}^{1}dx\int\frac{d^{D}k}{(2\pi)^{D}}\frac{p_{\rho}
p_{\gamma}p_{\sigma}p_{\delta}x^{2}(1 - x)^{2}}{[k^{2} + p^{2}x(1 - x)]^{2}}\\
 & = \frac{i}{15(4\pi)^{2}\epsilon}p_{\rho}p_{\sigma}p_{\gamma}p_{\delta} + \mathcal{O}(\epsilon
^{0})\\
I_{\rho\sigma\gamma\delta}^{(3)} & = \int_{0}^{1}dx\int\frac{d^{D}k}{(2\pi)^{D}}\frac{p_{\rho}
p_{\sigma}k_{\gamma}k_{\delta}x(x - 1)}{[k^{2} + p^{2}x(1 - x)]^{2}}\\
 & = \frac{i}{30(4\pi)^{2}\epsilon}p_{\rho}p_{\sigma}p^{2}\eta_{\gamma\delta} + \mathcal{O}
 (\epsilon^{0})\\
I_{\rho\sigma\gamma\delta}^{(4)} & = \int_{0}^{1}dx\int\frac{d^{D}k}{(2\pi)^{D}}\frac{k_{\rho}
k_{\sigma}p_{\gamma}p_{\delta}x(x - 1)}{[k^{2} + p^{2}x(1 - x)]^{2}}\\
 & = \frac{i}{30(4\pi)^{2}\epsilon}p_{\gamma}p_{\delta}p^{2}\eta_{\rho\sigma} + \mathcal{O}
 (\epsilon^{0})\\
I_{\rho\sigma\gamma\delta}^{(5)} & = \int_{0}^{1}dx\int\frac{d^{D}k}{(2\pi)^{D}}\frac{k_{\rho}
k_{\sigma}k_{\gamma}k_{\delta}}{[k^{2} + p^{2}x(1 - x)]^{2}}\\
 & = \frac{ip^{4}}{120(4\pi)^{2}\epsilon}[\eta_{\rho\gamma}\eta_{\sigma\delta} + \eta_{\rho
\delta}\eta_{\sigma\gamma} + \eta_{\rho\sigma}\eta_{\gamma\delta}] + \mathcal{O}(\epsilon^{0}).   
\end{split}
\end{equation}
We may now note the following results
\begin{equation}
\begin{split}
I_{\mu\nu\alpha\beta}^{(1)} & = - \mathcal{P}_{\mu\nu}^{ \ \ \rho\sigma}\mathcal{P}_{\alpha\beta}
^{ \ \ \gamma\delta}\eta_{\sigma\delta}p_{\rho}p_{\gamma}p^{2}\\
 & = - \frac{\kappa^{2}}{4}\Big[p_{\mu}p_{\alpha}\eta_{\nu\beta} + p_{\mu}p_{\beta}\eta_{\nu
\alpha} - p_{\mu}p_{\nu}\eta_{\alpha\beta}\\
 & + p_{\nu}p_{\alpha}\eta_{\mu\beta} + p_{\nu}p_{\beta}\eta_{\mu\alpha} - p_{\mu}p_{\nu}\eta
_{\alpha\beta}\\
 & - \eta_{\mu\nu}p_{\alpha}p_{\beta} - p_{\alpha}p_{\beta}\eta_{\mu\nu} + p^{2}\eta_{\mu\nu}
 \eta_{\alpha\beta}\Big]\\
\implies h^{\mu\nu}(-p) & I_{\mu\nu\alpha\beta}^{(1)}h^{\alpha\beta}(p) = 0,
\end{split}
\end{equation}
\begin{equation}
\begin{split}
I_{\mu\nu\alpha\beta}^{(2)} & = - \mathcal{P}_{\mu\nu}^{ \ \ \rho\sigma}\mathcal{P}_{\alpha\beta}
^{ \ \ \gamma\delta}p_{\rho}p_{\sigma}p_{\gamma}p_{\delta}\\
 & = - \frac{\kappa^{2}}{4}\Big[4p_{\mu}p_{\nu}p_{\alpha}p_{\beta} - 2\eta_{\mu\nu}p_{\alpha}
 p_{\beta}p^{2}\\
 & - 2\eta_{\alpha\beta}p_{\mu}p_{\nu}p^{2} + \eta_{\mu\nu}\eta_{\alpha\beta}p^{4}\Big]\\
\implies h^{\mu\nu}(-p) & I_{\mu\nu\alpha\beta}^{(2)}h^{\alpha\beta}(p) = 0,
\end{split}
\end{equation}
\begin{equation}
\begin{split}
I_{\mu\nu\alpha\beta}^{(3)} & = - \mathcal{P}_{\mu\nu}^{ \ \ \rho\sigma}\mathcal{P}_{\alpha\beta}
^{ \ \ \gamma\delta}p_{\rho}p_{\sigma}\eta_{\gamma\delta}p^{2}\\
 & = \frac{\kappa^{2}}{4}(D - 2)p^{2}\eta_{\alpha\beta}(2p_{\mu}p_{\nu} - p^{2}\eta_{\mu\nu})\\
\implies h^{\mu\nu}(-p) & I_{\mu\nu\alpha\beta}^{(3)}h^{\alpha\beta}(p) = 0,
\end{split}
\end{equation}
\begin{equation}
\begin{split}
I_{\mu\nu\alpha\beta}^{(4)} & = \frac{\kappa^{2}}{4}(D - 2)p^{2}\eta_{\mu\nu}(2p_{\alpha}
p_{\beta} - p^{2}\eta_{\alpha\beta})\\
\implies h^{\mu\nu}(-p) & I_{\mu\nu\alpha\beta}^{(4)}h^{\alpha\beta}(p) = 0,  
\end{split}
\end{equation}
\begin{equation}
\begin{split}
I_{\mu\nu\alpha\beta}^{(5)} & = - \mathcal{P}_{\mu\nu}^{ \ \ \rho\sigma}\mathcal{P}_{\alpha\beta}
^{ \ \ \gamma\delta}[\eta_{\rho\gamma}\eta_{\sigma\delta} + \eta_{\rho\delta}\eta_{\sigma\gamma}
+ \eta_{\rho\sigma}\eta_{\gamma\delta}]p^{4}\\
 & = - \frac{\kappa^{2}}{4}p^{4}[4\eta_{\mu\alpha}\eta_{\nu\beta} + 4\eta_{\mu\beta}\eta_{\nu
 \alpha} + (D^{2} - 2D - 4)\eta_{\mu\nu}\eta_{\alpha\beta}]\\
\implies & h^{\mu\nu}(-p)I_{\mu\nu\alpha\beta}^{(5)}h^{\alpha\beta}(p) = - \frac{\kappa^{2}}
{4}p^{4}[8h^{\alpha\beta}(-p)h_{\alpha\beta}(p)\\
 & + (D^{2} - 2D - 4)h(- p)h(p)].  
\end{split}
\end{equation}
\subsection{Dirac fields}
\label{App:1-LoopDirac}

Eq. (21) is the self-energy corrections to gravitons due to massless Dirac fields. After using the Feynman integral representation and doing some algebraic manipulations, 
we obtain the following integral expression
\begin{equation}
\begin{split}
\mathcal{J}^{\mu\nu\alpha\beta} & = i\mathcal{P}_{ \ \ \rho\sigma}^{\mu\nu}\mathcal{P}_{ \ \ 
\gamma\delta}^{\alpha\beta}\frac{\kappa^{2}}{4(4\pi)^{2}\epsilon}\int_{0}^{1}dx p^{2}x(1 - x)\\
 & \times\Big[(1 - 2x)^{2}p^{\rho}p^{\gamma}\eta^{\sigma\delta} - 2\eta^{\rho\gamma}
 p^{\sigma}p^{\delta}x(1 - x)\\
 & + \frac{1}{2}\eta^{\rho\sigma}p^{\gamma}p^{\delta}(1 - 2x)^{2} + \frac{1}{2}\eta^{\rho
 \delta}p^{\gamma}p^{\sigma}(1 - 2x)^{2}\\
 & + \frac{1}{2}\eta^{\gamma\sigma}p^{\rho}p^{\delta}(1 - 2x)^{2} + \frac{1}{2}\eta^{\gamma
 \delta}p^{\rho}p^{\sigma}(1 - 2x)^{2}\\
 & - \eta^{\sigma\delta}p^{\rho}p^{\gamma}(1 - 2x)^{2} + \eta^{\sigma\delta}\eta^{\rho\gamma}
 p^{2}x(1 - x)\\
 & - \frac{1}{2}\eta^{\sigma\delta}p^{\gamma}p^{\rho}(1 - 2x)^{2} - \frac{1}{2}\eta^{\sigma
 \delta}p^{\rho}p^{\gamma}(1 - 2x)^{2}\Big]\\
 & - i\mathcal{P}_{ \ \ \rho\sigma}^{\mu\nu}\mathcal{P}_{ \ \ \gamma\delta}^{\alpha\beta}\frac{\kappa^{2}}{8(4\pi)^{2}\epsilon}\int_{0}^{1}dx [\eta^{\sigma\delta}p^{\rho}p^{\gamma}
 p^{2}(1 - 2x)^{2}x(1 - x)\\
 & - 2p^{\rho}p^{\sigma}p^{\gamma}p^{\delta}(1 - 2x)^{2}x(1 - x)]\\
 & - i\mathcal{P}_{ \ \ \rho\sigma}^{\mu\nu}\mathcal{P}_{ \ \ \gamma\delta}^{\alpha\beta}\frac{3\kappa^{2}}{8(4\pi)^{2}\epsilon}\int_{0}^{1}dx \Big[\frac{1}{3}x^{2}(1 - x)^{2}\\
 & \times (\eta^{\rho\gamma}\eta^{\sigma\delta} + \eta^{\rho\sigma}\eta^{\gamma\delta}
 + \eta^{\rho\delta}\eta^{\sigma\gamma}) - \eta^{\sigma\delta}\eta^{\rho\gamma}x^{2}(1 - x)^{2}
 \Big]p^{4}\\
 & \equiv \mathcal{J}_{1}^{\mu\nu\alpha\beta} + \mathcal{J}_{2}^{\mu\nu\alpha\beta} + 
 \mathcal{J}_{3}^{\mu\nu\alpha\beta}.
\end{split}
\end{equation}
After evaluating the above set of integral, we obtain the following results
\begin{equation}
\begin{split}
h_{\mu\nu}(-p)\mathcal{J}_{1}^{\mu\nu\alpha\beta}h^{\alpha\beta}(p) & = \frac{i\kappa^{2}p^{4}}
{30(4\pi)^{2}\epsilon}[h_{\mu\nu}(- p)h^{\mu\nu}(p) - 2h(-p)h(p)]\\
h_{\mu\nu}(-p)\mathcal{J}_{2}^{\mu\nu\alpha\beta}h^{\alpha\beta}(p) & = - \frac{i\kappa^{2}p^{4}}
{60(4\pi)^{2}\epsilon}h(- p)h(p)\\
h_{\mu\nu}(-p)\mathcal{J}_{3}^{\mu\nu\alpha\beta}h^{\alpha\beta}(p) & = \frac{i\kappa^{2}p^{4}}
{60(4\pi)^{2}\epsilon}(h_{\mu\nu}(- p)h^{\mu\nu}(p) - h(- p)h(p)).
\end{split}
\end{equation}

\section{Linearised action of quadratic gravity theory}
\label{App:Linearised gravity QG}
In this section, we derive the linearised action corresponding to the general quadratic gravity theory (11) up to the quadratic order. We consider the metric perturbation around the
Minkowski spacetime following the previous section. Following the Appendix 
\ref{App:Linearised gravity}, we may write the following expression for the Riemann tensor
up to linear order
\begin{equation}
R_{ \ \nu\rho\sigma}^{\mu} = \bar{R}_{ \ \nu\rho\sigma}^{\mu} + \varepsilon(R_{ \ \nu\rho
\sigma}^{\mu})_{L} \equiv \varepsilon[\bar{\nabla}_{\rho}(\Gamma_{ \ \sigma\nu}^{\mu})_{L}
- \bar{\nabla}_{\sigma}(\Gamma_{ \ \rho\nu}^{\mu})_{L}],
\end{equation}
which is our case reduces to the following form
\begin{equation}
R_{\mu\nu\rho\sigma} = \frac{\varepsilon}{2}[\partial_{\rho}\partial_{\nu}h_{\sigma\mu} - 
\partial_{\rho}\partial_{\mu}h_{\sigma\nu} - \partial_{\sigma}\partial_{\nu}h_{\rho\mu}
+ \partial_{\sigma}\partial_{\mu}h_{\rho\nu}].
\end{equation}
As a result, up to the quadratic order, we obtain the following expression for the scalar
$R_{\mu\nu\rho\sigma}R^{\mu\nu\rho\sigma}$
\begin{equation}\label{Riemann squared}
\begin{split}
R_{\mu\nu\rho\sigma}R^{\mu\nu\rho\sigma} & = \varepsilon^{2}[h^{\mu\nu}(\Box)^{2}h_{\mu\nu}
 + 2\partial_{\sigma}h^{\sigma\mu}\Box\partial^{\rho}h_{\rho\mu}\\
 & + \partial_{\mu}\partial_{\nu}h^{\mu\nu}\partial^{\alpha}\partial^{\beta}h_{\alpha\beta}].
\end{split}
\end{equation}
On the other hand, the Ricci tensor at the linear order reduces to the following form
\begin{equation}
(R_{\nu\sigma})_{L} = \frac{\varepsilon}{2}(\partial_{\mu}\partial_{\sigma}h_{ \ \nu}^{\mu}
+ \partial_{\mu}\partial_{\nu}h_{ \ \sigma}^{\mu} - \Box h_{\sigma\nu} - \partial_{\sigma}
\partial_{\nu}h).
\end{equation}
In order to obtain the expression for $R_{\nu\sigma}R^{\nu\sigma}$ up to quadratic order, 
we make integration by parts after doing the contraction of indices which leads to the
following form
\begin{equation}\label{Ricci tensor squared}
\begin{split}
R_{\nu\sigma}R^{\nu\sigma} & = \frac{\varepsilon^{2}}{2}\Big[\partial_{\mu}\partial_{\nu}
h^{\mu\nu}\partial_{\rho}\partial_{\sigma}h^{\rho\sigma} - \partial_{\mu}\partial_{\nu}h
^{\mu\nu}\Box h\\
 & + \partial_{\mu}h^{\mu\nu}\Box\partial^{\sigma}h_{\sigma\nu} + \frac{1}{2}h_{\mu\nu}
 (\Box)^{2}h^{\mu\nu} + \frac{1}{2}h(\Box)^{2}h\Big].
\end{split}
\end{equation}
Further, the linearised expression of Ricci scalar is given by
\begin{equation}
\begin{split}
R_{L} & = \eta^{\nu\sigma}(R_{\nu\sigma})_{L} - h^{\nu\sigma}\bar{R}_{\nu\sigma} = 
\eta^{\nu\sigma}(R_{\nu\sigma})_{L}\\
 & = \varepsilon(\partial_{\mu}\partial_{\nu}h^{\mu\nu} - \Box h),
\end{split}
\end{equation}
and as a result, we obtain the following expression for $R^{2}$ up to quadratic order
\begin{equation}\label{Ricci squared}
R^{2} = \varepsilon^{2}[\partial_{\mu}\partial_{\nu}h^{\mu\nu}\partial_{\rho}\partial_{\sigma}
h^{\rho\sigma} - 2\partial_{\mu}\partial_{\nu}h^{\mu\nu}\Box h + h(\Box)^{2}h].
\end{equation}
Using the expressions in the equations \eqref{Riemann squared}, \eqref{Ricci tensor squared},
\eqref{Ricci squared}, we obtain the following relation
\begin{equation}
\begin{split}
a & R_{\mu\nu\rho\sigma}R^{\mu\nu\rho\sigma} + bR_{\mu\nu}R^{\mu\nu} + cR^{2}\\
 & = \varepsilon^{2}\Big[\left(a + \frac{b}{4}\right)h^{\mu\nu}(\Box)^{2}h_{\mu\nu} + 
 \left(\frac{b}{4} + c\right)h(\Box)^{2}h\\
 & + \left(a + \frac{b}{2} + c\right)\partial_{\mu}\partial_{\nu}h^{\mu\nu}\partial
_{\alpha}\partial_{\beta}h^{\alpha\beta} + \left(2a + \frac{b}{2}\right)\partial_{\sigma}
 h^{\sigma\mu}\Box\partial^{\rho}h_{\rho\mu}\\
 & - \left(\frac{b}{2} + 2c\right)\partial_{\mu}\partial_{\nu}h^{\mu\nu}\Box h\Big],
\end{split}
\end{equation}
which in the Lorenz gauge reduces to the following form
\begin{equation}
\begin{split}
& a  R_{\mu\nu\rho\sigma}R^{\mu\nu\rho\sigma} + bR_{\mu\nu}R^{\mu\nu} + cR^{2}\\
 & = \frac{\varepsilon^{2}}{4}[(4a + b)h^{\mu\nu}(\Box)^{2}h_{\mu\nu} + (c - a)h(\Box)^{2}h].
\end{split}
\end{equation}
We may note that in the case of GB gravity theory $b = - 4a, c = a$ which makes
the kinetic terms in the above action vanishing identically. This shows that there is no
propagating degrees of freedom in GB gravity in 4-D which makes it
a topological theory in 4-D.

\section{Non-minimal kinetic energy term through one-loop quantum correction}\label{Non-minimal term}
\begin{figure}
\includegraphics[height = 5cm, width = 5cm]{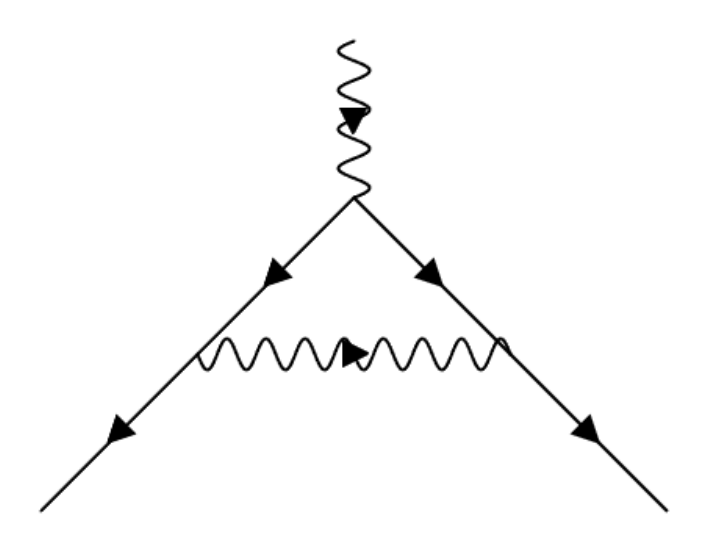}
\caption{Triangle diagram at one-loop level generating non-minimal kinetic energy term}
\label{Triangle diagram}
\end{figure}
In this section, we compute the triangle diagram as shown in Fig. \ref{Triangle diagram}
and show that it will eventually generate the non-minimal kinetic energy term. The integral
associated with this diagram can be expressed as
\begin{equation}
\begin{split}
\mathcal{I}_{\Delta} & = \int\frac{d^{D}k}{(2\pi)^{D}}\frac{1}{(p_{1} + k)^{2}k^{2}
(p_{2} - k)^{2}}\mathcal{V}_{\mu\nu}(p_{1} + p_{2}, p_{1} + k, p_{2} - k)\mathcal{V}
_{\rho\sigma}(k, - p_{1}, p_{1} + k)\\
 & \times\mathcal{V}_{\alpha\beta}(- k, - p_{2}, p_{2} - k)\mathcal{D}^{\rho\sigma
 \alpha\beta}\\
 & = \int\frac{d^{D}k}{(2\pi)^{D}}\frac{1}{(p_{1} + k)^{2}k^{2}(p_{2} - k)^{2}}
 \mathcal{P}_{\mu\nu}^{ \ \ \mu_{1}\nu_{1}}(p_{1} + k)_{\mu_{1}}(p_{2} - k)_{\nu_{1}}
 \mathcal{P}_{\rho\sigma}^{ \ \ \rho_{1}\sigma_{1}}(p_{1})_{\rho_{1}}(p_{1} + k)_{\sigma_{1}}
 \\
 & \times\mathcal{P}_{\alpha\beta}^{ \ \ \alpha_{1}\beta_{1}}(k - p_{2})_{\alpha_{1}}(p_{2})
 _{\beta_{1}}\times\Big[\frac{1}{2}(\eta^{\rho\alpha}\eta^{\sigma\beta} + \eta^{\rho\beta}
 \eta^{\sigma\alpha}) - \frac{1}{D - 2}\eta^{\rho\sigma}\eta^{\alpha\beta}\Big]\\
 & = - \frac{\kappa^{3}}{8}\int\frac{d^{D}k}{(2\pi)^{D}}\frac{1}{(p_{1} + k)^{2}k^{2}(p_{2}
 - k)^{2}}(p_{1} + k)_{\mu_{1}}(p_{2} - k)_{\nu_{1}}(p_{1})_{\rho_{1}}(p_{1} + k)_{\sigma_{1}}
 (k - p_{2})_{\alpha_{1}}(p_{2})_{\beta_{1}}\\
 & \times [\delta_{\mu}^{\mu_{1}}\delta_{\nu}^{\nu_{1}} + \delta_{\mu}^{\nu_{1}}\delta_{\nu}
 ^{\mu_{1}} - \eta_{\mu\nu}\eta^{\mu_{1}\nu_{1}}][\delta_{\rho}^{\rho_{1}}\delta_{\sigma}
 ^{\sigma_{1}} + \delta_{\rho}^{\sigma_{1}}\delta_{\sigma}^{\rho_{1}} - \eta_{\rho\sigma}
 \eta^{\rho_{1}\sigma_{1}}][\delta_{\alpha}^{\alpha_{1}}\delta_{\beta}^{\beta_{1}} + 
 \delta_{\alpha}^{\beta_{1}}\delta_{\beta}^{\alpha_{1}} - \eta_{\alpha\beta}\eta^{\alpha_{1}
 \beta_{1}}]\\
 & \times\Big[\frac{1}{2}(\eta^{\rho\alpha}\eta^{\sigma\beta} + \eta^{\rho\beta}\eta^{\sigma
 \alpha}) - \frac{1}{D - 2}\eta^{\rho\sigma}\eta^{\alpha\beta}\Big],  
\end{split}
\end{equation}
where
\begin{equation}
\begin{split}
[\delta_{\rho}^{\rho_{1}} & \delta_{\sigma}^{\sigma_{1}} + \delta_{\rho}^{\sigma_{1}}
\delta_{\sigma}^{\rho_{1}} - \eta_{\rho\sigma}\eta^{\rho_{1}\sigma_{1}}]\Big[\frac{1}{2}
(\eta^{\rho\alpha}\eta^{\sigma\beta} + \eta^{\rho\beta}\eta^{\sigma\alpha}) - \frac{1}{D - 2}
\eta^{\rho\sigma}\eta^{\alpha\beta}\Big][\delta_{\alpha}^{\alpha_{1}}\delta_{\beta}^{\beta_{1}} 
 + \delta_{\alpha}^{\beta_{1}}\delta_{\beta}^{\alpha_{1}} - \eta_{\alpha\beta}\eta^{\alpha_{1}
 \beta_{1}}]\\
 & \times (p_{1})_{\rho_{1}}(p_{1} + k)_{\sigma_{1}}(k - p_{2})_{\alpha_{1}}(p_{2})_{\beta_{1}}
 \\
 & = 2[2p_{1}.p_{2}(p_{1} - p_{2}).k - 2(p_{1}.p_{2})^{2} + k^{2}p_{1}.p_{2}].
\end{split}
\end{equation}
Therefore, after expressing the denominator in terms of Feynman parameter integral, we obtain
the following expression
\begin{equation}
\begin{split}
\mathcal{I}_{\Delta} & = - \frac{\kappa^{3}}{8}\int_{0}^{1}dx\int_{0}^{1}dy\int\frac{d^{D}k}{
(2\pi)^{D}} \ \Big[2p_{1}.p_{2}(p_{1})_{\mu_{1}}(p_{2})_{\nu_{1}}(p_{1} - p_{2}).k + 2p_{1}.p_{2}(p_{2})_{\nu_{1}}(p_{1} - p_{2}).k k_{\mu_{1}}\\
 & - 2p_{1}.p_{2}(p_{1})_{\mu_{1}}(p_{1} - p_{2}).k k_{\nu_{1}} - 2p_{1}.p_{2}(p_{1} - p_{2}).k
 k_{\mu_{1}}k_{\nu_{1}} - 2(p_{1}.p_{2})^{2}(p_{1})_{\mu_{1}}(p_{2})_{\nu_{1}} - 2(p_{1}.p_{2})^{2}
 (p_{2})_{\nu_{1}}k_{\mu_{1}}\\
 & + 2(p_{1}.p_{2})^{2}(p_{1})_{\mu_{1}}k_{\nu_{1}} + 2(p_{1}.p_{2})^{2}k_{\mu_{1}}k_{\nu_{1}}
 + k^{2}p_{1}.p_{2}(p_{1})_{\mu_{1}}(p_{2})_{\nu_{1}} + k^{2}p_{1}.p_{2}(p_{2})_{\nu_{1}}k_{\mu_{1}}
 - k^{2}p_{1}.p_{2}(p_{1})_{\mu_{1}}k_{\nu_{1}}\\
 & - k^{2}p_{1}.p_{2}k_{\mu_{1}}k_{\nu_{1}}\Big]\times\frac{1}{\Big[[k + (xp_{1} - yp_{2})]^{2}
 + 2xyp_{1}.p_{2}\Big]^{3}}\\
 & = \frac{\kappa^{3}}{8}\int\frac{d^{D}k}{(2\pi)^{D}}\left(\frac{1}{k^{2} + 2xyp_{1}.p_{2}}
 \right)^{3}\Big\{2(p_{1}.p_{2})^{2}(p_{1})_{\mu_{1}}(p_{2})_{\nu_{1}}(x + y) - 2(p_{1}.p_{2})^{2}
 (p_{2})_{\nu_{1}}(xp_{1} - yp_{2})_{\mu_{1}}\\
 & \times (x + y) + 2(p_{1}.p_{2})^{2}(p_{1})_{\mu_{1}}(xp_{1} - yp_{2})_{\nu_{1}}(x + y)
 - 2(p_{1}.p_{2})^{2}(x + y)(xp_{1} - yp_{2})_{\mu_{1}}(xp_{1} - yp_{2})_{\nu_{1}}\\
 & - 2(p_{1}.p_{2})^{2}(p_{1})_{\mu_{1}}(p_{2})_{\nu_{1}} + 2(p_{1}.p_{2})^{2}\{(xp_{1} - yp_{2})
 _{\mu_{1}}(p_{2})_{\nu_{1}} - (p_{1})_{\mu_{1}}(xp_{1} - yp_{2})_{\nu_{1}}\}\\
 & + 2(p_{1}.p_{2})^{2}(xp_{1} - yp_{2})_{\mu_{1}}(xp_{1} - yp_{2})_{\nu_{1}} - 2xy(p_{1}.p_{2})^{2}
 (p_{1})_{\mu_{1}}(p_{2})_{\nu_{1}} + 2xy(p_{1}.p_{2})^{2}(p_{2})_{\nu_{1}}(xp_{1} - yp_{2})_{\mu_{1}}
 \\
 & - 2xy(p_{1}.p_{2})^{2}(p_{1})_{\mu_{1}}(xp_{1} - yp_{2})_{\nu_{1}} + 2xy(p_{1}.p_{2})^{2}(xp_{1} 
 - yp_{2})_{\mu_{1}}(xp_{1} - yp_{2})_{\nu_{1}}\Big\}\\
 & \times [\delta_{\mu}^{\mu_{1}}\delta_{\nu}^{\nu_{1}} + \delta_{\mu}^{\nu_{1}}\delta_{\nu}^{\mu_{1}}
 - \eta_{\mu\nu}\eta^{\mu_{1}\nu_{1}}].
\end{split}
\end{equation}
In the $D = 4$ dimension, the above integral reduces to the following form
\begin{equation}
\mathcal{I}_{\Delta} = - \frac{i\kappa^{3}}{64\pi^{2}}(p_{1}.p_{2})\Big[\frac{2}{3}(p_{1})_{\mu}
(p_{1})_{\nu} + \frac{2}{3}(p_{2})_{\mu}(p_{2})_{\nu} - \frac{1}{2}(p_{1})_{\mu}(p_{2})_{\nu} - 
\frac{1}{2}(p_{2})_{\mu}(p_{1})_{\nu} + \frac{1}{2}\eta_{\mu\nu}p_{1}.p_{2}\Big].
\end{equation}
As a result, the one-loop corrected action also contains a term which is of the following form
\begin{equation}
\begin{split}
S_{\Delta} & = - \frac{\kappa^{3}}{64\pi^{2}}\int d^{4}x \Big[\frac{4}{3}h^{\mu\nu}\partial_{\mu}
\partial_{\nu}\partial_{\rho}\phi\partial^{\rho}\phi - h^{\mu\nu}\partial_{\mu}\partial_{\rho}\phi
\partial_{\nu}\partial^{\rho}\phi + \frac{1}{2}h\partial_{\mu}\partial_{\nu}\phi\partial^{\mu}
\partial^{\nu}\phi\Big].
\end{split}
\end{equation}
Since we have been using the de-Donder gauge fixing condition $\partial_{\mu}h^{\mu\nu} = \frac{1}{2}\partial^{\nu}h$ and do the perturbation computation along the tree-level theory where $\Box\phi = 0$ or in other words, $p^{2} = 0$ for on-shell four momenta, we finally obtain the following expression after using the integration by parts
\begin{equation}
S_{\Delta} = - \frac{7\kappa^{3}}{192\pi^{2}}\int d^{4}x \Big[\frac{1}{2}h\partial_{\mu}\partial_{\nu}
\phi\partial^{\mu}\partial^{\nu}\phi - h^{\mu\nu}\partial_{\mu}\partial_{\rho}\phi\partial_{\nu}\partial^{\rho}\phi\Big].
\end{equation}
On the other hand, the linearised Einstein tensor is expressed as
\begin{equation}
G^{\mu\nu} = \frac{\kappa}{2}\left(- \Box h^{\mu\nu} + \partial^{\mu}\partial_{\alpha}h^{\nu\alpha} + \partial^{\nu}\partial_{\alpha}h^{\mu\alpha} - \partial^{\mu}\partial^{\nu}h - \eta^{\mu\nu}\partial_{\alpha}\partial_{\beta}h^{\alpha\beta} + \eta^{\mu\nu}\Box h\right),
\end{equation}
which in de-Donder gauge reduces to the following form
\begin{equation}
G^{\mu\nu} = \frac{\kappa}{2}\left(- \Box h^{\mu\nu} + \frac{1}{2}\eta^{\mu\nu}\Box h\right).
\end{equation}
Using the above expression, we obtain the following relation
\begin{equation}
\begin{split}
G^{\mu\nu}\partial_{\mu}\phi\partial_{\nu}\phi & = \frac{\kappa}{2}\left( - \Box h^{\mu\nu}
\partial_{\mu}\phi\partial_{\nu}\phi + \frac{1}{2}\Box h \partial_{\mu}\phi\partial^{\mu}\phi\right),
\end{split}
\end{equation}
which reduces to the following form
\begin{equation}
\begin{split}
G^{\mu\nu}\partial_{\mu}\phi\partial_{\nu}\phi & = \kappa\Big[ - h^{\mu\nu}\Box\partial_{\mu}\phi
\partial_{\nu}\phi - h^{\mu\nu}\partial^{\rho}\partial_{\mu}\phi\partial_{\rho}\partial_{\nu}\phi
+ \frac{1}{2}h\Box\partial_{\mu}\phi\partial^{\mu}\phi + \frac{1}{2}h\partial_{\mu}\partial_{\rho}
\phi\partial^{\mu}\partial^{\rho}\phi\Big],
\end{split}
\end{equation}
after doing the integration by parts. Since we are considering perturbative approach, $\Box\phi = 0$
which leads to the following expression
\begin{equation}
G^{\mu\nu}\partial_{\mu}\phi\partial_{\nu}\phi = \kappa\left(\frac{1}{2}h\partial_{\mu}\partial_{\rho}
\phi\partial^{\mu}\partial^{\rho}\phi - h^{\mu\nu}\partial^{\rho}\partial_{\mu}\phi\partial_{\rho}\partial_{\nu}\phi\right).
\end{equation}
This shows that we may express the one-loop quantum corrected term $S_{\Delta}$ in effective action 
as
\begin{equation}
S_{\Delta} = - \frac{7\kappa^{2}}{192\pi^{2}}\int d^{4}x G^{\mu\nu}\partial_{\mu}\phi\partial_{\nu}
\phi.
\end{equation}
This clearly shows the generation of non-minimal kinetic coupling term in the one-loop quantum effective action.   

\section{Higher derivative term through one-loop quantum correction}
\begin{figure}
\includegraphics[height = 5cm, width = 5cm]{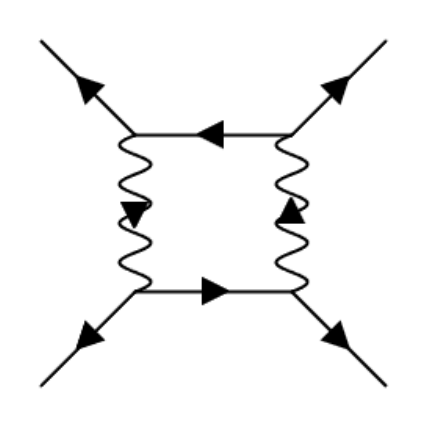}
\caption{Box diagram in one-loop quantum correction}
\label{Fig. 2}
\end{figure}

In this section, we compute the box diagram as shown in Fig. \ref{Fig. 2}.
This diagram can be computed through the computation of the following integral
\begin{equation}
\begin{split}
\mathcal{I}_{\Box} & = \int\frac{d^{D}k}{(2\pi)^{D}}\frac{1}{k^{2}(k + p_{1})^{2}(k + p_{1}
 + p_{2})^{2}(k + p_{1} + p_{2} + p_{3})^{2}}\mathcal{V}_{\mu\nu}(k, k + p_{1} + p_{2} + p_{3}, 
 - p_{1} - p_{2} - p_{3})\\
 & \times\mathcal{V}_{\rho\sigma}(- k, p_{1}, - k - p_{1})\mathcal{V}_{\gamma\delta}(k + p_{1}
 + p_{2}, k + p_{1}, p_{2})\mathcal{V}_{\alpha\beta}(- k - p_{1} - p_{2}, p_{3}, - k - p_{1}
 - p_{2} - p_{3})\\
 & \times\mathcal{D}^{\mu\nu\rho\sigma}\mathcal{D}^{\alpha\beta\gamma\delta}\\
 & = - \int\frac{d^{D}k}{(2\pi)^{D}}\frac{1}{k^{2}(k + p_{1})^{2}(k + p_{1} + p_{2})^{2}(k + p_{1} 
 + p_{2} + p_{3})^{2}}\mathcal{P}_{\mu\nu}^{ \ \ \mu_{1}\nu_{1}}(k + p_{1} + p_{2} + p_{3})_{\mu_{1}}
 (p_{1} + p_{2} + p_{3})_{\nu_{1}}\\
 & \times\mathcal{P}_{\rho\sigma}^{ \ \ \rho_{1}\sigma_{1}}(p_{1})_{\rho_{1}}(k + p)_{\sigma_{1}}
 \mathcal{P}_{\gamma\delta}^{ \ \ \gamma_{1}\delta_{1}}(k + p_{1})_{\gamma_{1}}(p_{2})_{\delta_{1}}
 \mathcal{P}_{\alpha\beta}^{ \ \ \alpha_{1}\beta_{1}}(p_{3})_{\alpha_{1}}(k + p_{1} + p_{2} + p_{3})
 _{\beta_{1}}\\
 & \times\Big[\frac{1}{2}(\eta^{\mu\rho}\eta^{\nu\sigma} + \eta^{\mu\sigma}\eta^{\nu\rho}) - 
 \frac{1}{D - 2}\eta^{\mu\nu}\eta^{\rho\sigma}\Big]\Big[\frac{1}{2}(\eta^{\alpha\gamma}\eta^{\beta
 \delta} + \eta^{\alpha\delta}\eta^{\beta\gamma}) - \frac{1}{D - 2}\eta^{\alpha\beta}\eta^{\gamma
 \delta}\Big]\\
 & = - \frac{\kappa^{4}}{16}\int\frac{d^{D}k}{(2\pi)^{D}}\frac{1}{k^{2}(k + p_{1})^{2}(k + p_{1} +  
 p_{2})^{2}(k + p_{1} + p_{2} + p_{3})^{2}}(k + p_{1} + p_{2} + p_{3})_{\mu_{1}}(p_{1} + p_{2} + 
 p_{3})_{\nu_{1}}\\
 & \times (p_{1})_{\rho_{1}}(k + p_{1})_{\sigma_{1}}(k + p_{1})_{\gamma_{1}}(p_{2})_{\delta_{1}}
 (p_{3})_{\alpha_{1}}(k + p_{1} + p_{2} + p_{3})_{\beta_{1}}[\delta_{\mu}^{\mu_{1}}\delta_{\nu}
 ^{\nu_{1}} + \delta_{\mu}^{\nu_{1}}\delta_{\nu}^{\mu_{1}} - \eta_{\mu\nu}\eta^{\mu_{1}\nu_{1}}]\\
 & \times\Big[\frac{1}{2}(\eta^{\mu\rho}\eta^{\nu\sigma} + \eta^{\mu\sigma}\eta^{\nu\rho}) - 
 \frac{1}{D - 2}\eta^{\mu\nu}\eta^{\rho\sigma}\Big][\delta_{\rho}^{\rho_{1}}\delta_{\sigma}
 ^{\sigma_{1}} + \delta_{\rho}^{\sigma_{1}}\delta_{\sigma}^{\rho_{1}} - \eta_{\rho\sigma}
 \eta^{\rho_{1}\sigma_{1}}][\delta_{\alpha}^{\alpha_{1}}\delta_{\beta}^{\beta_{1}} + \delta_{\alpha}
 ^{\beta_{1}}\delta_{\beta}^{\alpha_{1}} - \eta_{\alpha\beta}\eta^{\alpha_{1}\beta_{1}}]\\
 & \times\Big[\frac{1}{2}(\eta^{\alpha\gamma}\eta^{\beta\delta} + \eta^{\alpha\delta}\eta^{\beta
 \gamma}) - \frac{1}{D - 2}\eta^{\alpha\beta}\eta^{\gamma\delta}\Big][\delta_{\gamma}^{\gamma_{1}}
 \delta_{\delta}^{\delta_{1}} + \delta_{\gamma}^{\delta_{1}}\delta_{\delta}^{\gamma_{1}} - \delta
 _{\gamma\delta}\eta^{\gamma_{1}\delta_{1}}]. 
\end{split}
\end{equation}
In the above expression, we may write the following relation
\begin{equation}
\begin{split}
\frac{1}{k^{2}(k + p_{1})^{2}(k + p_{1} + p_{2})^{2}(k + p_{1} + p_{2} + p_{3})^{2}} & = 
\frac{1}{\Gamma(4)}\int_{0}^{1}d\xi_{1}\int_{0}^{1}d\xi_{2}\int_{0}^{1}d\xi_{3}\int_{0}^{1}
d\xi_{4} \frac{\delta(\xi_{1} + \xi_{2} + \xi_{3} + \xi_{4} - 1)}{\mathcal{D}^{4}},
\end{split}
\end{equation}
where
\begin{equation}
\begin{split}
\mathcal{D} & = [\xi_{1}k^{2} + \xi_{2}(k + p_{1})^{2} + \xi_{3}(k + p_{1} + p_{2})^{2} + \xi_{4}(k + p_{1} + p_{2} + p_{3})^{2}]\\
 & = [k + (\xi_{2} + \xi_{3} + \xi_{4})p_{1} + (\xi_{3} + \xi_{4})p_{2} + \xi_{4}p_{3}]^{2} + 2\xi_{1}
 \xi_{3}p_{1}.p_{3} + 2\xi_{2}\xi_{4}p_{2}.p_{3}.
\end{split}
\end{equation}
In order to obtain the above result, we have used the fact that $\xi_{1} + \xi_{2} + \xi_{3} + \xi_{4}
 = 1$ and $(p_{1} + p_{2} + p_{3})^{2} = 0$ which also implies $p_{1}.p_{3} = - (p_{1}.p_{2} + p_{2}.p_{3})$. This follows from the on-shell condition of the external legs in the box diagram. On the other hand, the numerator can be expressed as
\begin{equation}
\begin{split}
[\delta_{\mu}^{\mu_{1}}\delta_{\nu}^{\nu_{1}} + \delta_{\mu}^{\nu_{1}}\delta_{\nu}^{\mu_{1}} - \eta_{\mu\nu}\eta^{\mu_{1}\nu_{1}}] & \Big[\frac{1}{2}(\eta^{\mu\rho}\eta^{\nu\sigma} + \eta^{\mu\sigma}\eta^{\nu\rho}) - \frac{1}{D - 2}\eta^{\mu\nu}\eta^{\rho\sigma}\Big][\delta_{\rho}^{\rho_{1}}\delta_{\sigma}
 ^{\sigma_{1}} + \delta_{\rho}^{\sigma_{1}}\delta_{\sigma}^{\rho_{1}} - \eta_{\rho\sigma}
 \eta^{\rho_{1}\sigma_{1}}]\\
\times [\delta_{\alpha}^{\alpha_{1}}\delta_{\beta}^{\beta_{1}} + \delta_{\alpha}^{\beta_{1}}\delta_{\beta}^{\alpha_{1}} - \eta_{\alpha\beta}\eta^{\alpha_{1}\beta_{1}}] & \Big[\frac{1}{2}(\eta^{\alpha\gamma}\eta^{\beta\delta} + \eta^{\alpha\delta}\eta^{\beta\gamma}) - \frac{1}{D - 2}\eta^{\alpha\beta}\eta^{\gamma\delta}\Big][\delta_{\gamma}^{\gamma_{1}}\delta_{\delta}^{\delta_{1}} 
+ \delta_{\gamma}^{\delta_{1}}\delta_{\delta}^{\gamma_{1}} - \delta_{\gamma\delta}\eta^{\gamma_{1}\delta_{1}}]\\
= 4[\eta^{\mu_{1}\rho_{1}}\eta^{\nu_{1}\sigma_{1}} & + \eta^{\mu_{1}\sigma_{1}}\eta^{\nu_{1}\rho_{1}}
- \eta^{\mu_{1}\nu_{1}}\eta^{\rho_{1}\sigma_{1}}][\eta^{\alpha_{1}\gamma_{1}}\eta^{\beta_{1}\delta_{1}} + \eta^{\alpha_{1}\delta_{1}}\eta^{\beta_{1}\gamma_{1}} - \eta^{\alpha_{1}\beta_{1}}
\eta^{\gamma_{1}\delta_{1}}].
\end{split}
\end{equation}
As a result, we obtain the following expression
\begin{equation}
\begin{split}
\mathcal{I}_{\Box} & = - \frac{\kappa^{4}}{4}\int\frac{d^{D}k}{(2\pi)^{D}}\int_{0}^{1}d\xi_{1}
\int_{0}^{1}d\xi_{2}\int_{0}^{1}d\xi_{3}\int_{0}^{1}d\xi_{4}\frac{\delta(\xi_{1} + \xi_{2}
 + \xi_{3} + \xi_{4} - 1)}{\Gamma(4)\mathcal{D}^{4}}\\
 & [\eta^{\mu_{1}\rho_{1}}\eta^{\nu_{1}\sigma_{1}} + \eta^{\mu_{1}\sigma_{1}}\eta^{\nu_{1}\rho_{1}}
- \eta^{\mu_{1}\nu_{1}}\eta^{\rho_{1}\sigma_{1}}](k + p_{1} + p_{2} + p_{3})_{\mu_{1}}(p_{1} + p_{2} + p_{3})_{\nu_{1}}(p_{1})_{\rho_{1}}(k + p_{1})_{\sigma_{1}}\\
 & [\eta^{\alpha_{1}\gamma_{1}}\eta^{\beta_{1}\delta_{1}} + \eta^{\alpha_{1}\delta_{1}}\eta^{\beta_{1}\gamma_{1}} - \eta^{\alpha_{1}\beta_{1}}\eta^{\gamma_{1}\delta_{1}}](k + p_{1})_{\gamma_{1}}(p_{2})_{\delta_{1}}(p_{3})_{\alpha_{1}}(k + p_{1} + p_{2} + p_{3})_{\beta_{1}}.
\end{split}
\end{equation}
From the above expression, one can generate two quantum corrected term in the effective action which are $(\partial\phi)^{4}$ and $\phi^{2}(\partial\phi)^{2}$, however, we are interested in the first term. After doing the momentum integration, we obtain the following result
\begin{equation}
\begin{split}
\mathcal{I}_{\Box} & = i\frac{\kappa^{4}}{16(4\pi)^{2}}\int_{0}^{1}d\xi_{1}\int_{0}^{1}d\xi_{2}\int_{0}^{1}d\xi_{3}\int_{0}^{1}d\xi_{4} \frac{\delta(\xi_{1} + \xi_{2} + \xi_{3} + \xi_{4} - 1)}{(\xi_{1}\xi_{3}p_{1}.p_{2} + \xi_{2}\xi_{4}p_{2}.p_{3})^{2}}\Big[\{\xi_{3}p_{1}.p_{2} + (1 - \xi_{4})p_{2}.p_{3}\}\\
 & \times\{(1 - \xi_{2})p_{2}.p_{3} + \xi_{3}p_{1}.p_{2}\} + p_{2}.p_{3}\{2p_{1}.p_{2}\xi_{1}\xi_{3} 
 + p_{2}.p_{3}(2\xi_{2}\xi_{4} + 1 - \xi_{2} - \xi_{4})\}\\
 & + \{\xi_{2}p_{2}.p_{3} - \xi_{3}p_{1}.p_{2}\}\{\xi_{3}p_{1}.p_{2} - \xi_{4}p_{2}.p_{3}\}\Big]\Big[ - 
 \{\xi_{1}p_{1}.p_{2} + (1 - \xi_{2})p_{2}.p_{3}\}\{\xi_{1}p_{1}.p_{2} + (1 - \xi_{4})p_{2}.p_{3}\}\\
 & + p_{2}.p_{3}\{ - 2\xi_{1}\xi_{3}p_{1}.p_{2} + p_{2}.p_{3}((1 - \xi_{4})(\xi_{2} - 1) - \xi_{2}\xi_{4})\} + \{\xi_{1}p_{1}.p_{2} - \xi_{2}p_{2}.p_{3}\}\{\xi_{1}p_{1}.p_{2} - \xi_{4}p_{2}.p_{3}\}\Big]
 \\
 & = -i\frac{\kappa^{4}}{4(4\pi)^{2}}\int_{0}^{1}d\xi_{1}\int_{0}^{1}d\xi_{2}\int_{0}^{1}d\xi_{3}\int_{0}^{1}d\xi_{4} \frac{\delta(\xi_{1} + \xi_{2} + \xi_{3} + \xi_{4} - 1)}{(\xi_{1}\xi_{3}p_{1}.p_{2} + \xi_{2}\xi_{4}p_{2}.p_{3})^{2}} (p_{2}.p_{3})^{2}[p_{1}.p_{2}\xi_{1}\xi_{3} + (p_{2}.p_{3})\xi_{2}\xi_{4}\\
 & + p_{1}.p_{2}\xi_{3} + p_{2}.p_{3}(1 - \xi_{2} - \xi_{4})][(p_{1}.p_{2})\xi_{1}\xi_{3} + (p_{2}.p_{3})\xi_{2}\xi_{4} + p_{1}.p_{2}\xi_{1} + p_{2}.p_{3}(1 - \xi_{2} - \xi_{4})]. 
\end{split}
\end{equation}
In order to compute the contribution coming from the term $(\partial\phi)^{4}$, we consider the following integral contribution
\begin{equation}
\begin{split}
\bar{\mathcal{I}}_{\Box} & = - i\frac{\kappa^{4}}{4(4\pi)^{2}}(p_{2}.p_{3})^{2}\int_{0}^{1}d\xi_{1}\int_{0}^{1}d\xi_{2}\int_{0}^{1}d\xi_{3}\int_{0}^{1}d\xi_{4}\delta(\xi_{1} + \xi_{2} + \xi_{3} + \xi_{4} - 1)\\
 & = - i\frac{\kappa^{4}}{4(4\pi)^{2}}(p_{2}.p_{3})^{2}\int_{0}^{1}d\xi_{1}\int_{0}^{1 - \xi_{1}}d\xi_{2} (1 - \xi_{1} - \xi_{2}) =  - i\frac{\kappa^{4}}{24(4\pi)^{2}}(p_{2}.p_{3})^{2}.
\end{split}
\end{equation}
As a result, we obtain the following term in the quantum effective action in four dimensions
\begin{equation}
\begin{split}
S_{\Box} & = - \frac{\kappa^{4}}{64\pi^{2}}\int d^{4}x \ \phi^{2}\partial_{\mu}\partial_{\nu}
\phi\partial^{\mu}\partial^{\nu}\phi\\
 & = \frac{\kappa^{4}}{64\pi^{2}}\int d^{4}x \ 2\phi\partial_{\nu}\phi\partial_{\mu}\phi
 \partial^{\mu}\partial^{\nu}\phi\\
 & = - \frac{\kappa^{4}}{64\pi^{2}}\int d^{4}x \ (\partial_{\mu}\phi\partial^{\mu}\phi)^{2},
\end{split}
\end{equation}
where we neglected the terms involving $\Box\phi$ since it vanishes as we consider the perturbative
corrections.

\bibliographystyle{apsrev}
\bibliography{Ver-Fin}

\end{document}